\documentclass[12pt]{article}
\usepackage{latexsym}
\usepackage[utf8]{inputenc}
\usepackage[T1]{fontenc}
\usepackage[normalem]{ulem} 
\usepackage{cite,epsfig,amssymb,euscript,slashed}
\usepackage{amsmath}
\usepackage{array,calc,epsfig}
\usepackage{braket}
\usepackage{xcolor}
\usepackage{float}
\usepackage{bbm}
\usepackage{fancybox}

\def\be{\begin{equation}}
\def\ee{\end{equation}}
\def\bseq{\begin{subequations}}
\def\eseq{\end{subequations}}

\def\bea{\begin{eqnarray}}
\def\eea{\end{eqnarray}}
\newcommand\bbone{\ensuremath{\mathbbm{1}}}
\newcommand{\ul}{\underline}
\def\bseq{\begin{subequations}}
\def\eseq{\end{subequations}}

\numberwithin{equation}{section} 
\usepackage[]{graphicx}

\def\d {{\rm d}}

\def\cala         {{\cal A}}

\def\calb         {{\cal B}}
\def\calc         {{\cal C}}
\def\cald         {{\cal D}}
\def\cale         {{\cal E}}
\def\calf         {{\cal F}}
\def\calg         {{\cal G}}

\def\calh         {{\cal H}}
\def\cali         {{\cal I}}

\def\calk         {{\cal K}}
\def\call         {{\cal L}}
\def\calm         {{\cal M}}
\def\caln         {{\cal N}}

\def\calp         {{\cal P}}

\def\calr         {{\cal R}}
\def\cals         {{\cal S}}
\def\calt         {{\cal T}}

\def\calw         {{\cal W}}
\def\calz         {{\cal Z}}

\def\iund{\underline{i}}

\def\ibar{{\overline{\imath}}}

\def\del          {\partial}

\def\ii           {{\rm i}}

\def\Re           {{\rm Re\hskip0.1em}}
\def\Im           {{\rm Im\hskip0.1em}}

\def\half{{\frac12}}

\def\sqr#1#2{{\vcenter{\vbox{\hrule height.#2pt
 \hbox{\vrule width.#2pt height#1pt \kern#1pt \vrule width.#2pt}\hrule
 height.#2pt}}}}

\def\d{\text{d}}

\def\slashchar#1{\setbox0=\hbox{$#1$}           
\dimen0=\wd0                                 
\setbox1=\hbox{/} \dimen1=\wd1               
\ifdim\dimen0>\dimen1                        
\rlap{\hbox to \dimen0{\hfil/\hfil}}      
#1                                        
\else                                        
\rlap{\hbox to \dimen1{\hfil$#1$\hfil}}   
/                                         
\fi}

\begin{document}
\font\cmss=cmss10 \font\cmsss=cmss10 at 7pt

\title{\vspace{-2.0cm}\begin{flushright}{\scriptsize DFPD-2018/TH/01}
\end{flushright}
\hfill\\
\hfill\\
Three-forms, dualities and membranes  
\\[0.1cm] 
in 
four-dimensional 
 supergravity
\\[0.8cm] }

\author{Igor Bandos$^{\dagger\ddagger}$, Fotis Farakos$^\spadesuit$, Stefano Lanza$^{\ast\diamondsuit}$,\\ Luca Martucci$^{\ast\diamondsuit}$ and Dmitri Sorokin$^{\diamondsuit\ast}$}

\date{}

\maketitle

\vspace{-1.5cm}

\begin{center}

\vspace{0.5cm}
\textit{\small $^{\dagger}$ Department of
Theoretical Physics, University of the Basque Country UPV/EHU,
P.O. Box 644, 48080 Bilbao, Spain
 \\
$^{\ddagger}$ IKERBASQUE, Basque Foundation for Science, 48011, Bilbao, Spain \\
{} $^\spadesuit$ KU Leuven, Institute for Theoretical Physics,  Celestijnenlaan 200D,\\ B-3001 Leuven, Belgium 
\\
$^{\ast}$ Dipartimento di Fisica e Astronomia ``Galileo Galilei",  Universit\`a degli Studi di Padova,\\
Via F. Marzolo 8, 35131 Padova, Italy
\\ 
$^\diamondsuit$ I.N.F.N. Sezione di Padova, Via F. Marzolo 8, 35131 Padova, Italy}
\end{center}

\vspace{5pt}

\abstract{
\noindent We consider four-dimensional $\caln=1$ supergravity models of a kind appearing in string flux compactifications.
It has been recently shown that,  by using double three-form multiplets instead of ordinary chiral multiplets, one can promote to dynamical variables (part of) the quantized numbers appearing in the flux-induced   superpotential. 
We show that double three-form multiplets  naturally transform under symplectic dualities associated with the special K\"ahler structure that characterizes their scalar sector. 
Furthermore, we discuss how to couple membranes which carry arbitrary `electric-magnetic' charges. 
The complete  action is  supersymmetric,  kappa-symmetric 
and duality covariant. As an application, we derive the flow equations for BPS  domain walls sourced by  membranes and give simple analytic examples of their  solution.}


\thispagestyle{empty}


\newpage

\setcounter{footnote}{0}

\tableofcontents

\newpage

\section{Introduction}
In recent years the role of gauge three-forms in four-dimensional effective theories has been revisited in various contexts and from different perspectives. In particular, three-forms can provide an effective dynamical `Hodge-dual' description of the internal flux quanta which specify certain classes of string flux compactifications. The inclusion  of gauge three-forms in effective theories may have interesting physical effects, such as the dynamical generation of the cosmological constant and its neutralization \cite{Duff:1980qv,Aurilia:1980xj,Hawking:1984hk,Brown:1987dd,Brown:1988kg,Duff:1989ah,Duncan:1989ug,Ovrut:1997ur,Bousso:2000xa,Feng:2000if,Wu:2007ht,Bandos:2010yy,Bandos:2011fw,Bandos:2012gz,Farakos:2016hly}, the contribution to inflationary scenarios \cite{Kaloper:2008fb,Kaloper:2011jz,Marchesano:2014mla,Bielleman:2015ina,Dudas:2014pva,Valenzuela:2016yny}, possible resolution of the strong CP violation problem \cite{Dvali:2005an,Dvali:2004tma,Dvali:2005zk,Dvali:2013cpa,Dvali:2016uhn,Dvali:2016eay} and others.

In \cite{Farakos:2017jme} it was shown how to construct generic  ${\mathcal N}=1$ supergravity models in the presence of gauge three-forms.  
The construction  naturally applies to supergravities with a scalar sector described by a special K\"ahler geometry. This is indeed often the case for stringy effective theories, as the ones  considered  in  \cite{Bielleman:2015ina,Carta:2016ynn, Herraez:2018vae}, which studied the role of gauge three-forms in the context of type II    compactifications. The link between double three-forms and the moduli parametrizing  a special K\"ahler space can be understood, for instance, by thinking of type IIB theory on a Calabi--Yau three-fold and focusing on the special K\"ahler space parametrized by its complex structure moduli. As usual, one can consider a symplectic basis of internal CY three-cycles    $(\Pi^I,\tilde \Pi_J)$ ($I,J=0,\ldots,n$) and parametrize the complex structures moduli by  projective coordinates $s^I\equiv \int_{\Pi^I}\Omega$, where $\Omega$ is the holomorphic CY $(3,0)$ form\cite{Candelas:1990pi}.  On the other hand, by integrating the Ramond-Ramond (R-R) six-forms on $(\Pi^I,\tilde \Pi_J)$ one gets a set of associated double three-forms $(A^I_3,\tilde A_{3I})$ in four-dimensions.  Now,  this and similar settings are characterized by a symplectic duality group associated with the special K\"ahler geometry, which in this example corresponds just to the group of possible symplectic rotations of the basis $(\Pi^I,\tilde \Pi_J)$.
These transformations mix up the coordinates $s^I$ with the dual coordinates $\calg_I(s)\equiv \int_{\tilde\Pi_I}\Omega$. At the same time, they should  act on the double three-forms $(A^I_3,\tilde A_{3I})$ as well, rotating them as electric-magnetic pairs.

These observations may be extrapolated to other less supersymmetric examples, with orientifolds and internal fluxes, whose effective theory can be described by an $\caln=1$ supergravity with double three-form multiplets  of the form derived in \cite{Farakos:2017jme}, or a variation thereof. The double three-form multiplets contain a complex scalar and a pair of three-forms, which will be denoted by $s^I$ and $(A^I_3,\tilde A_{3J})$ as in  the above example. We then expect  these supergravities 
to naturally accommodate the action of the symplectic group associated with the relevant special K\"ahler structure.  In this paper we show that this is indeed the case: the symplectic  group has a natural action on the double three-form multiplets and the effective supergravity is covariant under it. 

Since our supersymmetric formulation contains three-form gauge potentials, it is also quite natural to consider the corresponding charged objects, namely membranes sweeping three-dimensional world-volumes $\calc$. By requiring compatibility with the symplectic transformations, they should couple to the three-forms through a bosonic minimal-coupling term of the form
\be\label{bWZ}
q_I\int_\calc A^I_{3}-p^I\int_\calc \tilde A_{3I} \, , 
\ee
where $(p^I,q_J)$ denote a symplectic vector of quantized electro-magnetic charges, which rotate under the action of the symplectic duality group. 
For instance, in the above type IIB example, a $(p^I,q_J)$-membrane would correspond to a D5-brane wrapping an internal three-cycle $\Pi$ homologous to $q_I\Pi^I-p^I\tilde\Pi_I$. 

Such membranes can be coupled in a manifestly supersymmetric way to the bulk sector of our four-dimensional effective supergravity. Indeed, we will show how to construct an action for a supermembrane in a curved superspace which couples to the bulk three-forms as in \eqref{bWZ}. This action contains a Nambu--Goto (NG) and a Wess--Zumino (WZ) term and is $\kappa$-symmetric, generalizing previously constructed supermembrane actions in four dimensions \cite{Achucarro:1988qb,Ovrut:1997ur,Huebscher:2009bp,Bandos:2010yy,Bandos:2011fw,Bandos:2012gz,Kuzenko:2017vil} and analogous superstring actions with two-form couplings in $D=3$ \cite{Buchbinder:2017qls}. The WZ term, which is the supersymmetric completion of  \eqref{bWZ}, will be defined by constructing appropriate super-three-forms from the three-form multiplets. On the other hand, the NG term will be uniquely fixed by requiring that the membrane action is $\kappa$-symmetric and hence supports a supersymmetric spectrum. The result is that the tension $T_M$ of the $(p^I,q_J)$-membrane is not a constant but rather depends on the scalar sector of the theory, namely
\be\label{tension}
T_M=2\left|q_I s^I-p^I\calg_I(s)\right| \, . 
\ee
Note that,  in the above type IIB example, this formula precisely matches the effective tension of the wrapped D5-branes, which  is given by the volume of the internal cycle  $\Pi$. Indeed, the BPS condition corresponds to a calibration condition \cite{harvey1982} which implies that  $T_M=\left|\int_{\Pi}\Omega\right|=\big|q_I\int_{\Pi^I}\Omega-p^I\int_{\tilde\Pi_I}\Omega\big|$,  hence reproducing    \eqref{tension}. This observation can be easily extended  to more general and less supersymmetric type II flux compactifications on generalized Calabi-Yau spaces \cite{Martucci:2005ht,Martucci:2006ij}. 

We will provide explicit examples of $\mathcal N=1$  supergravity models coupled to  double three-form multiplets and supermembranes. These include the effective theory resulting from type IIA  compactifications with R-R fluxes.  We stress that, differently from what happens for strings and branes in  String Theory and  M-theory, in four dimensions the membrane $\kappa$-symmetry  does not require the bulk sector to obey any dynamical equations. To be precise, $\kappa$-symmetry imposes  a set of {\em off-shell} constraints on the curved target superspace torsion which does not result in any equations of motion for the gravity or matter multiplets. This allows us to consider interacting systems in which the supermembrane back-reacts on the dynamical `bulk' supergravity sector. Particular examples of such systems, namely a supermembrane interacting with a single three-form matter multiplet and with single three-form supergravity were proposed and studied in \cite{Bandos:2010yy} and \cite{Bandos:2011fw,Bandos:2012gz}, respectively. 

In this paper we discuss in detail the general structure of BPS domain wall configurations arising in the supergravity theories under consideration, which
significantly enlarge the class of domain walls previously described in an ordinary chiral matter-coupled supergravity \cite{Cvetic:1992bf, Cvetic:1992st, Cvetic:1992sf, Cvetic:1993xe,Cvetic:1996vr, Ceresole:2006iq} and in a  three-form supergravity \cite{Ovrut:1997ur,Huebscher:2009bp}. 

We think that the actions constructed in this paper provide an appropriate framework for describing, from an effective four-dimensional perspective, non-trivial dynamical processes involving at the same time membranes, fluxes and the scalar sector of flux compactifications, as for instance those considered in \cite{Bousso:2000xa,Feng:2000if}. 

The paper is organized as follows. In Section \ref{sec:Sugra} we review one of the main results of  \cite{Farakos:2017jme}, showing how to trade a chiral matter-coupled $\caln=1$ supergravity  for a dual theory that contains double three-form multiplets. In Section \ref{sec:subclass} we specialize to the case where the kinetic terms for the double three-form multiplets are defined by an underlying special K\"ahler geometry. These theories are naturally covariant under special K\"ahler symplectic tranformations, under which the  gauge three-forms rotate as symplectic vectors. We also discuss the quantization of vacuum expectation values of four-form fluxes in purely four-dimensional setting. 
In Section \ref{sec:+membrane} we study the coupling \eqref{bWZ} of the gauge (super) three-forms to supersymmetric membranes. The requirement of the supermembrane action to be invariant under local $\kappa$-symmetry fixes its Nambu-Goto term such that the corresponding tension has the form \eqref{tension}.

With all the ingredients settled, in Section \ref{sec:jumpingDW} we consider the complete bulk-plus-membrane action  and study 
BPS domain walls interpolating between different supersymmetric vacua separated by the membrane. The vevs of the four-form fluxes are discontinuous (i.e. `jump') across the membrane. When the three-forms are integrated out, they give rise to two different (disconnected) effective scalar field superpotentials on each side of the membrane. We will see that, as expected, for BPS domain walls the tension \eqref{tension} precisely balances the  minimal-coupling term \eqref{bWZ} which cancel out in the complete bulk-plus-membrane BPS action, whose on-shell value depends only on the asymptotic vacua.

In Sections \ref{sec:example} and \ref{sec:exDWs} we study a simple model involving only two double three-form multiples, hence containing four three-form gauge potentials.
This model has a single vacuum for each choice of the four-form field-strength integration constants, and domain walls interpolate between the two different vacua separated by the membrane. We explicitly construct analytic solutions describing these domain walls.

In the Appendices, we collect additional results. In particular, in Appendix \ref{app:kappa} we provide  the complete proof of $\kappa$-symmetry for the supermembrane action and in Appendix \ref{app:gener} we describe the extension of this result to supermembranes coupled to more general bulk sectors including single three-form multiples.


\section{Supergravity with double three-forms}
\label{sec:Sugra}

In order to make this paper self-contained and to set the notation, in this section we briefly review the structure of the $\caln=1$ supergravity with three-form multiplets constructed in \cite{Farakos:2017jme}.  We will focus on the locally supersymmetric case with double three-form multiplets, but the construction under consideration can be easily adapted to the rigid case and/or to the other kinds of three-form multiplets studied e.g. in
\cite{Gates:1980ay,Gates:1980az,Buchbinder:1988tj,Ovrut:1997ur,Binetruy:2000zx,Bandos:2010yy,Bandos:2011fw,Bandos:2012gz,Farakos:2017jme,Kuzenko:2017vil,Buchbinder:2017vnb,Farakos:2017ocw}. 
Some of such examples are described in Appendix \ref{app:gener}.

Consider a general $\caln=1$ supergravity theory describing the coupling of the gravity multiplet to a set of chiral multiplets $(Z^I,T^r)$, with $I=0,\ldots,n$ and $r=1,\ldots,m$.  As in \cite{Farakos:2017jme}, we start from the super-conformally invariant approach \cite{Howe:1978km,Siegel:1978fc,Kugo:1982mr} (see e.g. \cite{Buchbinder:1995uq} for details) by including in $Z^I$ a conformal compensator such that the scaling dimension of $Z^I$ is $\Delta_Z=3$, while $T^r$ are assumed to have vanishing scaling dimension, $\Delta_T=0$.  Furthermore, we assume that $Z^I$ can be regarded as projective coordinates of a special K\"ahler manifold locally specified by a homogeneous prepotential $\calg(Z)$ such that $\calg(\lambda Z)=\lambda^2\calg(Z)$, for any arbitrary chiral superfield $\lambda$. We can now write the supergravity Lagrangian in the form
\be\label{WIactio}
\call'=-3\int\d^4\theta\,E\,\Omega(Z,\bar Z;T,\bar T) + \left( \int\d^2\Theta\,2\cale\,\calw(Z;T) + c.c. \right)\,.
\ee
Super-Weyl invariance  requires  that the `kinetic' superfield   $\Omega(Z,\bar Z;T,\bar T)$ has scaling dimension $\Delta_\Omega=2$, while the superpotential  $\calw(Z;T)$  has scaling dimension $\Delta_\calw=3$, 
which means 
\be
\begin{aligned}
\Omega(\lambda Z,\bar\lambda\bar Z;T,\bar T)=|\lambda|^{\frac23}\Omega(Z,\bar Z;T,\bar T)\,,\quad\calw(\lambda Z,T)=\lambda\calw( Z,T) . 
\end{aligned}
\ee
 Furthermore, we split $\calw(Z;T)$ as follows
\be\label{complsup}
\calw(Z;T)=e_IZ^I-m^I\calg_I(Z)+\hat \calw( Z;T) , 
\ee
where $\calg_I(Z)\equiv\del_I\calg(Z)=\calg_{IJ}Z^J$, $\calg_{IJ}\equiv \del_I\del_J\calg(Z)$, etc., and $\hat \calw( \lambda Z;T)=\lambda\hat \calw( Z;T)$. 
It is instructive to notice that the homogeneity requirements imply 
\be\label{cGIJ}
\calg_I(Z)= \calg_{IJ}(Z)Z^J\; , \qquad  \calg_{IJK}(Z)Z^K=0\; .\quad 
\ee
    
At this point, one may fix the super-Weyl invariance and get a standard supergravity action. However,  
by preserving super-Weyl invariance one can more easily derive from \eqref{WIactio} a dual theory in which the constants $(m^I,e_J)$ are promoted to the (Hodge-dual of the) field strengths of gauge three-forms. In short (we refer to \cite{Farakos:2017jme} for details), to get the dual Lagrangian one removes the terms $e_IZ^I-m^I\calg_I(Z)$ from the superpotential \eqref{complsup} and substitutes $Z^I$ with special chiral superfields $S^I$ constructed as follows
\be\label{Sdef}
S^I\equiv \frac14(\bar\cald^2-8\calr)\calm^{IJ}(\Sigma_J-\bar\Sigma_J).
\ee
Here $\calm^{IJ}$ is the inverse of 
\be
\calm_{IJ}\equiv \Im\calg_{IJ}
\ee
and $\Sigma_I$ are complex linear superfields, i.e. they satisfy  
\be\label{CLcond}
(\bar\cald^2-8\calr)\Sigma_I=0.
\ee
Note that  $\calg_{IJ}$ and $\calm_{IJ}$, which were functions of $Z^I$ and $\bar{Z}{}^I$, should now be considered as functions of $S^I$ and $\bar S^I$.  The new Lagrangian takes the form
\be\label{3formlagr}
\call=-3\int\d^4\theta\,E\,\Omega(S,\bar S;T,\bar T)+\left(\int\d^2\Theta\,2\cale\,\hat\calw(S;T)+ c.c. \right)+\call_{\rm bd},
\ee
where $\call_{\rm bd}$ is an appropriate boundary term which is necessary for having a well defined variational problem \cite{Farakos:2017jme}.

The complex linear superfields $\Sigma_I$ contain the double three-form multiplets, while the chiral superfields $S^I$ can be interpreted as multiplets of the corresponding field-strengths. Indeed, they are invariant under the gauge transformations
\be\label{gauge3form}
\Sigma_I\quad\rightarrow\quad \Sigma_I+\tilde L_I-\calg_{IJ}L^J \, , 
\ee
where $(L^I,\tilde L_J)$ are arbitrary real linear superfield parameters which supersymmetrize the three-form gauge transformations. 
We may partially fix (\ref{gauge3form}) by imposing a Wess-Zumino gauge in which, in particular, $\Sigma_I|=0$. The remaining bosonic degrees of freedom contained in  $\Sigma_I$  are  given by complex scalars ${\tt s}_I$ and double (real) three-forms $(A^I_3,\tilde A_{3J})$, which appear in the components\footnote{Here we use the convention $({}^*\omega_p)_{m_{p+1} \ldots m_4}=-\frac{\sqrt{-\det g}}{p!(4-p)!}\epsilon_{m_1\ldots m_4  }\omega^{m_1\ldots m_p}$ for any bosonic $p$-form $\omega_p$, where $\epsilon^{0123}=-\epsilon_{0123}=1$.
}
\be\label{compSigma}
\begin{aligned}
\cald^2\Sigma_I|&=-4\bar {\tt s}_I,\\
\bar\sigma_m^{\dot\alpha\alpha}[\cald_\alpha,\bar\cald_{\dot\alpha}]\Sigma_I\big|&=-2\big({}^*\!\tilde A_{3I}-\calg_{IJ}{}^*\! A_3^J\big)_m.
\end{aligned}
\ee
The residual gauge freedom in (\ref{gauge3form}) corresponds to  the standard gauge transformations $A^I_3\rightarrow A^I_3+\d\Lambda^I_2$  and $\tilde A_{3J}\rightarrow \tilde A_{3J}+\d\tilde\Lambda_{2J}$,  where $(\Lambda^I_2,\tilde\Lambda_{2J})$ are arbitrary two-forms. 

The lowest components  of the chiral superfields $S^I$ \eqref{Sdef} are related to
\eqref{compSigma} as follows
\be
S^I| \equiv s^I=\calm^{IJ}(s,\bar s) {\mathtt{s}}_J.
\ee
On the other hand, if we put to zero the fermions, the field-strengths $F^I_4\equiv \d A^I_3$ and $\tilde F_{4J}\equiv\d\tilde A_{3J}$  enter the $S^I$ highest component as follows
\be\label{SF}
F^I_S\equiv -\frac14\cald^2 S^I|=\bar M s^I-\frac\ii2\,\calm^{IJ}{}^*\!\calf_{4J}\,,
\ee
where $M$ is the complex scalar auxiliary field of the supergravity multiplet
and 
\be\label{defcalf}
\calf_{4I}\equiv \tilde F_{4I}-\bar\calg_{IJ} F_4^J\,
\ee
are complex four-forms with $\tilde F_{4I}=\d\tilde A_{3I}$ and $F_4^J=\d A_3^J$.

It is straightforward to extract from \eqref{3formlagr} the component Lagrangian and fix the super-Weyl invariance, 
finally obtaining, in the Einstein frame, supergravity with double three-forms, scalars and their fermionic partners. 
In the following sections we will restrict to a particular subclass  of models. 


\section{A special subclass of models}
\label{sec:subclass}

Let us now assume that  the kinetic function $\Omega$  has the factorized structure
\be\label{Omega}
\Omega(S,\bar S;T,\bar T)=\Omega_0(S,\bar S) e^{-\frac13 \hat K(T,\bar T)} \, ,
\ee
where  $\Omega_0(S,\bar S)$ is defined by the special K\"ahler structure as follows
\be\label{Omega0}
\Omega_0(S,\bar S) =\left[\ii \bar S^I\calg_I(S)-\ii S^I\bar\calg_I(\bar S) \right]^\frac{1}{3} \, . 
\ee
Furthermore, in order to simplify the presentation,  we will take $\hat\calw(S,T)\equiv 0$, so that the superpotential of the original theory is just given by $\calw(Z)=e_IZ^I-m^I\calg_I(Z)$. A non-trivial  $\hat\calw(S,T)$ may be easily incorporated by using the general results of Appendix \ref{app:boslagr}. The corresponding theory with the double three-form multiplets is then described by the Lagrangian
\be\label{3formlagr2}
\call=-3\int\d^4\theta\,E\,\Omega_0(S,\bar S)e^{-\frac13 \hat K(T,\bar T)}+\call_{\rm bd} \, . 
\ee
Having in mind the symplectic invariance discussed in the following section, we will gauge fix super-Weyl symmetry  in a slightly more flexible  way than in \cite{Farakos:2017jme}.

Let us first express $S^I$ in terms of new chiral superfields $Y$ and $\Phi^i$, with $i=1,\ldots,n$, as follows
\be\label{SYPHI}
S^I=Y\, f^I(\Phi) \, , 
\ee
where  $f^I(\Phi)$ are holomorphic functions of $\Phi^i$ such that $\text{rank}(\del_if^I)=n$. We  assume that the new fields have scaling dimensions $\Delta_Y=3$ and $\Delta_\Phi=0$, so that $Y$ can be regarded  as the compensator. $Y$ and $\Phi^i$ are not generic chiral superfields but rather have a constrained form, which can be in principle obtained by expressing them in terms of $S^I$ and then using \eqref{Sdef}. 

We can then write 
\be
\Omega_0(S,\bar S)=|Y|^\frac{2}{3}e^{-\frac13 K(\Phi,\bar\Phi)} \, , 
\ee
where 
\be
\label{SKpot}
\begin{aligned}
 K(\Phi,\bar\Phi)&=-\log\left[\ii \bar f^I(\bar\Phi)\calg_I(\Phi)-\ii  f^I(\Phi)\bar\calg_I(\bar\Phi)\right]\\
 &=-\log\left[-2\calm_{IJ}(\Phi,\bar\Phi) f^I(\Phi) \bar f^J(\bar\Phi)\right] \, , 
 \end{aligned}
\ee
with $\calg_I(\Phi)\equiv \calg_I(f(\Phi))$. We can now fix the super-Weyl gauge symmetry by imposing, for instance, 
\be\label{Wgfix}
Y=1 \, , 
\ee
so that the Lagrangian becomes\footnote{We work with Plank units $M^2_{\rm P}=1$.}
\be\label{3formlagr3}
\call=-3\int\d^4\theta\,E\,e^{-\frac13 \calk(\Phi,\bar\Phi,T,\bar T)}+\call_{\rm bd} \, , 
\ee
where 
\be\label{compK}
\calk(\Phi,\bar\Phi,T,\bar T)\equiv  K(\Phi,\bar\Phi)+ \hat K(T,\bar T) \, .
\ee
Of course there is a freedom in the choice of \eqref{SYPHI}. This can be associated with the possibility of redefining    $Y\rightarrow e^{g(\Phi)}Y$ and $f^I(\Phi)\rightarrow e^{-g(\Phi)}f^I(\Phi)$, which corresponds to a  K\"ahler transformation
\be\label{kahlertr}
K\rightarrow K+g(\Phi)+\bar g(\bar\Phi) \, .
\ee

\subsection{Bosonic action}
\label{sec:bosaction}
In order to express this Lagrangian in components, one should take into account \eqref{Sdef} and \eqref{Wgfix}. 
For simplicity, let us focus on the bosonic sector, setting  all fermions to zero and writing $Y=y+\Theta^2 F_Y$ and $\Phi^i=\phi^i+\Theta^2 F^i_\Phi$. From \eqref{SYPHI} it follows that
\be
F^I_S=f^I(\phi)F_Y+y f^I{}_i(\phi) F^i_\Phi , 
\ee
where $f^I{}_i(\phi)\equiv\frac{\del }{\partial \phi^i} f^I(\phi)\equiv\del_if^I(\phi)$ \footnote{In general, for any function of $\phi$ (and $\bar\phi$), e.g. $K(\phi,\bar\phi)$ we define $K_i\equiv\frac{K}{\partial \phi^i}\equiv\del_iK$, $K_{\bar i}\equiv\frac{K}{\partial \bar\phi^{\bar i}}\equiv\del_{\bar i}K$ etc.}. Then (\ref{Wgfix}) is equivalent to $y=1$, i.e.\ $s^I=f^I(\phi)$,
and $F^I_S= f^I{}_i F^i_\Phi$. In turn, by recalling \eqref{SF}, the latter is equivalent to
\be\label{FF}
 f^I(\phi)\bar M -f^I{}_i(\phi) F^i_\Phi= \frac\ii2\,\calm^{IJ}{}^*\!\calf_{4J}.
\ee
Since we are assuming that the change of coordinates  \eqref{SYPHI} is well defined, the $(n+1)\times (n+1)$ matrix $(f^I, f^I{}_i)$ is invertible and then \eqref{FF} can be inverted to express $M$ and $F^i_\Phi$ in terms of the field-strengths $(F^I_4,\tilde F_{4J})$. Note that $\calm^{IJ}$ and $\calg_{JK}$ should now be considered as functions of $(\phi^i,\bar\phi^{\bar\imath}$). 

Upon expanding \eqref{3formlagr3} in bosonic components,  performing the usual Weyl rescaling 
\be
\label{WeylResc}
 e^a_m \to e^a_m e^{\frac {1}{6} \calk}\,,\qquad (F^i_\Phi,F^q_T, M)  \to e^{-\frac 23 \calk}(F^i_\Phi,F^q_T, M)
\ee
to pass to the Einstein frame, integrating out $F^q_T$ and using \eqref{FF}, we arrive at the bosonic action 
\be\label{boscomp}
\begin{aligned}
S_{\rm bos}=&- \int \d^4 x\, e \Big(\frac{1}{2}R + G_{IJ} f^I{}_i \bar{f}^{J}{}_{\bar\jmath}\; \del \phi^i \del \bar{\phi}^{\bar \jmath}  + \hat K_{p \bar q}\,\del t^p \del\bar t^{\bar q}-\calt^{IJ}{}^*{}\!\bar\calf_{4I} {}^*{}\!\calf_{4J}\Big)+S_{\rm bd}  
\end{aligned}
\ee
with the boundary term\footnote{For a moment we are omitting the standard Gibbons-Hawking boundary term \cite{Gibbons:1976ue}, which will however be needed below, when we discuss domain wall solutions.} 
\be\label{bdaction}
\begin{aligned}
S_{\rm bd} 
&=\, 2\Re\int_\calb \calt^{IJ}(\tilde A_{3I}-\calg_{IK}A_3^K)\,{}^*\!\calf_{4J}\\
&=-2\Re\int \d^4 x\, e\,\nabla_m \left[ \calt^{IJ}\big( {}^*\!\tilde A_{3I}-\calg_{IK}{}^*\!A_3^K\big)^m\,{}^*\!\calf_{4J}\right], 
\end{aligned}
\ee
where $\calb$ is a space-time boundary (at infinity). In the above action, we have introduced the following quantities
\begin{subequations}
\begin{align}
G_{IJ}&\equiv-\frac{\calm_{IJ}}{(f \calm \bar f)}+\frac{(\calm\bar f)_I(\calm f)_{J} }{(f \calm \bar f)^2},\label{GIJ}\\ 
\calt^{IJ}&\equiv \frac14\, e^{-\calk}\left[\calm^{IK}G_{LK}\calm^{LJ}+\frac{1}{\gamma}\frac{f^I\bar f^J}{(f\calm\bar f)^2}\right],\\
\gamma &\equiv \hat{K}_{\bar q} \hat{K}^{\bar q p} \hat{K}_{p}-3\label{noscale}
\end{align}
\end{subequations}
with $(\calm f)_I\equiv \calm_{IJ}f^J$, $(\calm \bar f)_I\equiv \calm_{IJ}\bar f^J$ and $( f \calm \bar f)\equiv f^I \calm_{IJ}\bar f^J$. 
The inverse matrix of $\calt^{IJ}$ defined by $\calt_{IJ}\calt^{JK}=\calt^{KJ}\calt_{JI}=\delta_I^K$ has the following form
\be\label{invcalt}
\calt_{IJ}=-4e^\calk \left[(f\calm \bar f)\calm_{IJ}-(1+\gamma)(\calm f)_I(\calm \bar f)_J\right]\,.
\ee

Clearly, $\gamma$ must be non-vanishing in order for the above action to make sense. Indeed,  in deriving \eqref{boscomp} a vanishing  $\gamma$ would imply an obstruction in integrating out the auxiliary fields of the `spectator' chiral multiplets $T^p$.  In the following, we will assume that $\gamma$ is  non-vanishing and constant. For instance, in the case of type II orientifold compactifications one finds $\gamma=1$ \cite{Grimm:2004ua,Grimm:2004uq}. Or, in the absence of a spectator sector, we have $\gamma=-3$.

Note that $\overline {G_{IJ}}=G_{JI}$,   $f^IG_{IJ}=G_{JI}\bar f^I\equiv 0$ and $\overline{\calt^{IJ}}=\calt^{JI}$. Furthermore, 
the special K\"ahler structure requires that $\calm_{IJ}f^I\bar f^J<0$ and that the kinetic matrix  $G_{IJ}f^I{}_i\bar f^J{}_{\bar\jmath}$ is positive definite, see for instance \cite{Craps:1997gp}. Note also that 
\be\label{Kinv}
K^{\bar k l}\bar f^K_{\bar k}f^L_l G_{IK}G_{LJ}=G_{IJ}.
\ee
This can be verified by projecting the first and second index along the complete bases $(f^I,f^I_i)$ and  $(\bar f^J,\bar f^J_{\bar\jmath})$, respectively, and recalling that $K_{i\bar\jmath}=G_{IJ}f^I_i\bar f^J_{\bar\jmath}$.  

 The three-form equations of motion are
 \be\label{fluxeq}
 \d\,\Re(\calt^{IJ}{}^*\!\calf_{4J})=0\,,\quad  \d\,\Re(\calg_{IJ}\calt^{JK}{}^*\!\calf_{4K})=0\,. 
\ee
These equations imply that $\Re(\calt^{IJ}{}^*\!\calf_{4J})$ and $\Re(\calg_{IJ}\calt^{JK}{}^*\!\calf_{4K})$ are constant, at least away from the membrane sources introduced below, namely
\be\label{incost}
2\Re(\calt^{IJ}{}^*\!\calf_{4J})=m^I\,,\qquad 2\Re(\calg_{IJ}\calt^{JK}{}^*\!\calf_{4K})=e_I,
\ee
or equivalently 
\be\label{calfalphabeta}
\calt^{IJ}{}^*\!\calf_{4J}=-\frac\ii2\calm^{IJ}(e_J-\bar\calg_{JK}m^K),
\ee
where $m^I,e_J$ are real constants.

Consistent boundary conditions require the same combinations of the four-forms and scalars to take the fixed constant value on the boundary $\calb$. One can then check that, indeed, the boundary term \eqref{bdaction} makes the variational principle well defined, see for instance  \cite{Brown:1988kg, Groh:2012tf} for a discussion of this issue in simpler settings.

\subsection{Duality to standard matter-coupled supergravity} 
\label{sec:equiv}

Let us now explicitly check the relation of the above formulation to the ordinary bosonic supergravity action (in the absence of membranes). First, we notice that the part of the action \eqref{boscomp} containing the three-forms can be written in the following form
\be\label{3Fa}
\begin{aligned}
 S_{\text{3-forms}}&=\int \d^4 x\,e\, \calt^{IJ}{}^*\!\bar\calf_{4I}\, {}^*\!\calf_{4J}+S_{\rm bd}\\
& =
 -\int\d^4 x\,e\, \calt^{IJ}{}^*\!\bar\calf_{4I}\, {}^*\!\calf_{4J}\\
 &\quad-2\int \Big[\tilde A_{3I}\wedge\d\,\Re(\calt^{IJ}{}^*\!\calf_{4J})-A^I_{3}\wedge\d\,\Re(\calg_{IJ}\calt^{JK}{}^*\!\calf_{4K})\Big].
\end{aligned}
\ee
For further comparison with the conventional supergravity action let us define the following quantities 
\begin{subequations}
\label{W=}
\begin{align}
W_I&\equiv-2\ii \calm_{IJ}\calt^{KJ}{}^*\!\bar\calf_{4K}, \\
W&\equiv f^IW_I=\frac{\ii}{\gamma}e^{-\hat K}\left(f^I{}^*\tilde F_{4I}-\calg_I{}^* F^I_{4}\right),\\
W_i&\equiv f_i^IW_I\,.
\end{align}
\end{subequations}
At the moment this is  just a change of variables, but we will shortly see that $W$ is associated with the superpotential of the conventional supergravity.

By using the relations \eqref{GIJ}-\eqref{Kinv} it can be  checked that the scalar potential, 
which is given by 
\be\label{F4=V}
V\equiv  \calt^{IJ}{}^*\!\bar\calf_{4I}\, {}^*\!\calf_{4J} 
\ee
can be more explicitly written as follows 
\be\label{potst1}
\begin{split}
V
&= e^\calk \left[(f\calm \bar f)^2 \calm^{IK}\calm^{JL} G_{LK} W_I\overline W_{\bar J}+\gamma|W|^2\right]\\
&=e^\calk\left( K^{i\bar\jmath}(W_i+K_iW)( \overline W_{ \bar\jmath}+K_{\bar\jmath}\overline W)+\gamma|W|^2\right).
\end{split}
\ee
Now, if the four-forms satisfy the equations \eqref{fluxeq}-\eqref{calfalphabeta}, the action \eqref{3Fa} reduces to
\be\label{potential}
\begin{split}
S_{\text{3-forms}}&= -\int \d^4 x\,e\,\calt^{IJ}{}^*\!\bar\calf_{4I}\, {}^*\!\calf_{4J} \\
&\equiv -\int\d^4 x\sqrt{-g}\, V(\phi,\bar\phi,t,\bar t;e,m)\,, \quad~~~\text{(on-shell 3-forms)}
\end{split}
\ee
which corresponds to an effective potential $V$ for the scalar fields, in which however the coupling parameters $(e_I,m^I)$ are generated dynamically by the expectation values of the four-form fluxes.
Notice that the contribution of the non-vanishing boundary term \eqref{bdaction} has been crucial for getting the correct potential.

In view of \eqref{calfalphabeta}, the quantities $W$ and $W_i$ defined in \eqref{W=} become the following holomorphic functions of the scalar fields $\phi^i$
\be\label{Esup}
\begin{aligned}
W_I&= e_I-m^J\calg_{JI}(\phi),\\
W&=  e_If^I(\phi)-m^I\calg_I(\phi),\\
W_i&=\partial_iW,
\end{aligned}
\ee
and \eqref{potst1} reduces to
\be\label{potst}
\begin{aligned}
V&=e^\calk\left( K^{i\bar\jmath}D_iW\bar D_{\bar\jmath}\overline W+\gamma|W|^2\right) ,
\end{aligned}
\ee
where $D_iW(\phi)\equiv(\partial_i+K_i)W$ and $K_i=\partial_i K= - \frac {f_i\calm \bar{f}}{f\calm \bar{f}}$ with $K$ from (\ref{SKpot}).

Therefore, when the three-forms are integrated out, $W(\phi)$ is identified with the superpotential of the standard Einstein supergravity formulation obtained by gauge-fixing to the Einstein frame  the original Weyl invariant Lagrangian \eqref{WIactio} with $\calw=e_IZ^I-m^I\calg_I(Z)$. 

This discussion shows how the general superspace arguments of \cite{Farakos:2017jme} work in the bosonic sector of the theory.

It is worth mentioning that the supergravity models studied in this paper do include effective theories originating from flux compactifications of Type IIA and IIB string theory. Indeed, the  superpotential \eqref{Esup} is of the same form as that obtained in \cite{Gukov:1999ya, Taylor:1999ii}, where the constants $e_I, m^J$ are ultimately interpreted as quanta of background fluxes. In \cite{Farakos:2017jme} the three-form potential \eqref{potst1} was explicitly computed for a case of Type IIA effective theories, matching, on-shell, with the well known results from flux compactifications \cite{Louis:2002ny,Grimm:2004ua}. Owing to the generality of the previous discussion, the results obtained in this Section can also be extended to describe a landscape of Type IIB flux compactifications with orientifolds (see e.g. \cite{Blumenhagen:2003vr,Grimm:2004uq,Lust:2005bd}). 

In the context of effective theories arising from string compactifications, setting the gauge three-forms on-shell as in \eqref{incost} amounts to choosing a particular configuration of internal fluxes. However, we emphasize that in our three-form formulation the internal flux quanta are promoted to  unfixed dynamical quantities and, as we will see in the following sections, one can naturally include membranes, which can mediate dynamical transitions between different choices of flux quanta.  Our formulation
then provides a description of the effective theories originating from flux compactifications in which it is
possible to access the landscape of all the vacua corresponding to all different choices
of fluxes and the possible transitions between them within a single four-dimensional
theory.

\subsection{Quantization of integration constants}
\label{sec:quant}

In the above discussion, the integration constants $(m^I,e_J)$ introduced in \eqref{incost} and appearing in the effective superpotential \eqref{Esup} have been treated as arbitrary real constants. However, this is really so if the gauge three-forms $(A^I_3,\tilde A_{3J})$ are associated to  non-compact gauge symmetries. On the other hand, constructions from string theory, as well as purely four-dimensional arguments (see for instance \cite{Banks:2010zn}), indicate that in consistent quantum gravitational theories all gauge symmetries should be compact.  
In practice, this means that the integrals
\be\label{varthetab}
\varphi^I\equiv \int_\cale A^I_3\quad,\quad  \tilde\varphi_J\equiv\int_\cale \tilde A_{3J}
\ee
over any compact three-dimensional submanifold $\cale$, are periodic. It is then natural to normalize the gauge three-form fields so that \eqref{varthetab} have  $2\pi$-periodicity.

The compactness of the gauge symmetries implies quantization conditions on the corresponding field strengths. As in \cite{Kaloper:2011jz}, a simple way to identify these conditions is to relate our system to a  $1$-dimensional theory by performing the dimensional reduction  of the four-dimensional theory $\mathbb{R}\times \cale$ along the compact space-like $\cale$.  Let us focus on the four-form part of the bosonic action \eqref{boscomp}, namely 
\be
\int_{\mathbb{R}} \d t\, L_{\rm flux}\equiv-\int_{\mathbb{R}\times\cale}\calt^{IJ}\bar\calf_{4I}\, {}^*\!\calf_{4J},
\ee
where $t$ parametrizes the time direction $\mathbb{R}$.
In the  $1$-dimensional effective theory one can  compute the momenta conjugated to the angles $(\varphi^I,\tilde\varphi_J)$, 
namely 
\be\label{momenta}
\begin{aligned}
 \pi^I&\equiv \frac{\del L_{\rm flux}}{\del\dot{\tilde\varphi}_I}=-2\Re\left(\calt^{IJ}{}^*\!\calf_{4J}\right),\\
\tilde\pi_I&\equiv \frac{\del L_{\rm flux}}{\del\dot\varphi^I}=2\Re\left(\calg_{IJ}\calt^{JK}{}^*\!\calf_{4K}\right), 
\end{aligned}
\ee
where $\dot\varphi^I,\dot{\tilde\varphi}_J$ denote the  derivatives of the angles with respect to $t$. Quantum mechanically, the momenta must be integrally quantized: 
$\pi^I,\tilde\pi_J\in\mathbb{Z}$, since the angles are $2\pi$-periodic.
On the other hand, by comparing \eqref{momenta} and \eqref{incost} we see that  $-\pi^I$ and $\tilde\pi_J$ coincide  with the integration constants $m^I$ and $e_J$ respectively. Hence, we arrive at the quantization condition     
\be\label{bulkquant}
 e_I,m^J \in\mathbb{Z} \, . 
\ee
 This shows how  the compactness of the gauge symmetries  implies the quantization of the constants appearing in the effective superpotential \eqref{Esup}.
This is in agreement with what is expected from explicit string theory models, in which  $e_I,m^J$ usually measure quantized internal fluxes. However, we stress that  the three-form formulation has allowed for a
purely four-dimensional derivation of this fact.

In string models the quantized constants $e_I,m^J$ may need to satisfy an additional tadpole cancellation condition, which would fix the value of a linear combination thereof. In our formulation with three-forms, implementing this constraint would require to integrate out a single real gauge three-form which is a particular linear combination of the original $2n+2$ ones and to select a specific value of the corresponding integration constant. Thus the supergravity effective theory with the remaining $2n+1$ independent three-forms will identically satisfy the tadpole cancellation condition. For  simplicity, in this  paper we do not further consider this possibility.


\section{Symplectic covariance}
\label{sec:Sympl}

The above models of supergravities with gauge three-forms is based on the existence of a local special geometry defined by the prepotential $\calg(S)$, see for instance \cite{Freedman:2012zz} for a recent review on special geometry and more references. In fact, one can formulate the models without actually using the prepotential $\calg(S)$ but rather the $2(n+1)$-dimensional vector 
  \be\label{symplV}
V\equiv \left(\begin{array}{c}
 S^I \\ \calg_J
 \end{array}\right).
 \ee
One can immediately recognize that our general formulation of the double three-form multiplets is covariant under the symplectic transformations
 \be\label{Vsympl}
 V\quad\mapsto \quad \hat V=\cals V,
 \ee
where $\cals$ is an  Sp$(2n+2;\mathbb{R})$ matrix, i.e.\ such that $\cals^T\cali\cals=\cali$ and
 \be\label{symplmatrix}
\cali=\left(\begin{array}{rr} 0 & -\bbone \\
 \bbone & 0\end{array}\right).
 \ee 
 We will say that a $(2n+2)$-dimensional vector transforms as a symplectic vector if it  transforms as $V$ in \eqref{Vsympl}. 
 We can write  
 \be\label{defcalS}
\cals=\left(\begin{array}{cc} A & B \\
 C & D\end{array}\right),
 \ee
where  $A,B,C,D$ are $n\times n$ constant matrices such that
\be
A^TD-C^TB=\bbone\,,\quad A^TC-C^TA=0\,,\quad B^TD-D^TB=0.
\ee
Notice that, assuming that $A+B\mathbb{G}$ is invertible,  \eqref{Vsympl} is equivalent to
\begin{subequations}
\begin{align}
\hat S^I&= (A+B\mathbb{G})^I{}_JS^J\label{deltaSI},\\
\hat{\mathbb{G}}&=(C+D\mathbb{G})(A+B\mathbb{G})^{-1}\label{deltaG}, 
\end{align}
\end{subequations}
where 
$\mathbb{G}\equiv (\calg_{IJ})$. Furthermore, $\calm_{IJ}$ transforms as follows 
\be 
\hat\calm=(A+B\bar{\mathbb{G}})^{-1 T}\calm(A+B\mathbb{G})^{-1 }\label{deltaM}.
\ee

In order to better understand the action of the  above symplectic transformations on the elementary degrees of freedom of the double three-form multiplets, we observe that \eqref{deltaSI} can be alternatively defined by
\be\label{dualSigma}
\hat\Sigma_I= \Sigma_J[(A+B \mathbb{G})^{-1}]^J{}_I \, , 
\ee
as can be readily checked by using \eqref{Sdef}.  

One may in fact remove the condition on the non-degeneracy of $A+B\mathbb{G}$  by introducing the `prepotentials' $(\calp^I,\tilde\calp_J)$ defined as follows
\be\label{realprepot1}
\calp^I\equiv -2 \calm^{IJ}\Im\Sigma_J\quad,\quad \tilde\calp_I\equiv -2\Im(\bar\calg_{IJ}\calm^{JK}\Sigma_K) \, , 
\ee
which are such that 
\be\label{defSG}
S^I=-\frac\ii4(\bar\cald^2-8\calr)\calp^I\,,\quad \calg_I=-\frac\ii4(\bar\cald^2-8\calr)\tilde\calp_I \, . 
\ee 
Hence 
\be\label{calpvec}
\left(\begin{array}{c}
\calp^I \\ \tilde\calp_{J}
 \end{array}\right)
\ee
transforms as a symplectic vector and encodes, in a symplectic covariant way, all the  degrees of freedom of the double three-form multiplets $\Sigma_I$. 
This indicates that the supergravity with double-three forms may be formulated directly in terms of the symplectic vectors, without requiring the existence of a symplectic prepotential ${\cal G}(S)$, but just assuming ${\cal G}_I={\cal G}_{IJ}(S)S^J$. 
However, we will not attempt a complete discussion of such an intrinsic formulation and for simplicity will assume the existence of a prepotential, which is always available in an appropriate duality frame \cite{Craps:1997gp}.  

From these observations  one can  extract how the double three-forms transform. If all fermions vanish, combining \eqref{dualSigma} and \eqref{compSigma} we find that  the $(2n+2)$-dimensional vectors
\be
\left(\begin{array}{c}
A^I_3 \\ \tilde A_{3J}
 \end{array}\right)
\ee
transform as symplectic vectors.

Note that the extension of covariance to the original theory (with the ordinary chiral multiplets $Z^I$) we started from requires that the constants $(m^I,e_J)$ must transform as a symplectic vector. In this way, the form of the superpotential $e_I Z^I-m^I\calg_I$ is preserved.  On the other hand, the quantizations conditions discussed in Section \ref{sec:quant} imply that  the Sp$(2n+2;\mathbb{R})$ group is reduced to  a discrete subgroup Sp$(2n+2;\mathbb{Z})$. This structure naturally appears in stringy effective theories.

So far we have described the actions of the above  transformations on the symplectic vectors  \eqref{symplV} and  \eqref{calpvec}  of the super-Weyl invariant formulation. On the other hand, in order to fix the super-Weyl gauge invariance, one should express  $S^I$ (or, in a more intrinsic formulation which does not assume the prepotential $\calg$, all the components of the symplectic vector  \eqref{symplV}) as local holomorphic  functions of the `inhomogeneous coordinates' $(Y,\Phi^i)$ as in \eqref{SYPHI}. By identifying $\calg_I(f(\Phi))\rightarrow \calg_I(\Phi)$, a symplectic transformation maps the symplectic vector
\be\label{symplV2}
V(Y,\Phi)= Y\left(\begin{array}{c}
 f^I(\Phi) \\ \calg_J(\Phi)
 \end{array}\right) 
 \ee
to a new symplectic vector depending holomorphically on $(Y,\Phi^i)$. These symplectic transformations generically relate different equivalent choices of $(f^I(\Phi),\calg_J(\Phi))$ and guarantee the symplectic covariance, but not necessarily the invariance, of the theory. 
On the other hand, a  duality  symmetry of the special K\"ahler structure is a  transformation of $(Y,\Phi^i)$ that induces  a symplectic transformation of \eqref{symplV2}. For a simple example see Section \ref{sec:example}.
\footnote{In particular, by homogeneity, $\Phi^i$ will be mapped to new $\hat\Phi^i$ while $Y$ will be transformed into  $\hat Y=Y e^{g(\hat\Phi)}$, for some  holomorphic function $g(\hat\Phi)$. 
One can use $g(\Phi)$ to preserve the gauge fixing-condition \eqref{Wgfix} by combining the symplectic transformation with the field redefinition of the kind discussed just after  \eqref{compK}, which corresponds to a K\"ahler transformation.} 

Notice that the above discussion has not involved the kinetic potential $\Omega$, which should be appropriately restricted to be symplectic-invariant.  This does happen for the class of models considered in Section \ref{sec:subclass}.
Using the bosonic components of these transformations, one can explicitly check that the form of the bosonic Lagrangian \eqref{boscomp} is left invariant by the symplectic transformations.


\section{Inclusion of membranes}
\label{sec:+membrane}

In this section we include supersymmetric membranes with charges $(q_I,p^J)$ minimally coupled to the double three-form multiplets $(A^I_3,\tilde A_{3 J})$. In order to keep the supersymmetry manifest,  we promote the bosonic embedding of the membrane world-volume  $\calc$ to an embedding into the $\caln=1$ superspace extension of  four-dimensional space-time which is defined by the map
\be\label{superembedding}
\xi^i\quad\mapsto\quad z^M(\xi)=\left(x^m(\xi),\theta^\mu(\xi),\bar\theta^{\dot\mu}(\xi)\right) , 
\ee
where the $\xi^i$ with $i=0,1,2$, are the membrane world-volume coordinates. 
The bosonic minimal-coupling term \eqref{bWZ} can then be supersymmetrized to
\be
\label{susyWZ} 
S_{\rm WZ}=\int_{{\cal C}}{\cal A}_3
\ee 
with
\be\label{calA3=A+tA}
\cala_{3}= q_I \cala_{3}^I - p^I \tilde{\cala}_{3I} , 
\ee 
once we provide a set of appropriate {\em super} three-forms $(\cala_3^I,\tilde\cala_{3I})$ whose lowest components coincide with $(A_3^I,\tilde A_{3I})$. These super three-forms are constructed in terms of the pairs of the real `prepotentials' $(\calp^I,\tilde\calp_J)$ \eqref{realprepot1}. Given a prepotential $\calp$, the associated super three-form $\cala_3$ is defined by 
\be\label{super3form}
\begin{aligned}
\cala_{3}=&\,  { -}2 {\ii} E^a \wedge E^\alpha \wedge \bar E^{\dot\alpha}  \sigma_{a\alpha\dot\alpha}\calp + {\frac 12}  E^b\wedge E^a \wedge  E^\alpha
\sigma_{ab\; \alpha}{}^{\beta}{\cald}_{\beta}\calp \\ &+{\frac 12}  E^b\wedge E^a \wedge  \bar E^{\dot\alpha}
\bar\sigma_{ab}{}^{\dot\beta}{}_{\dot\alpha}\bar{\cald}_{\dot\beta}\calp  
\\&+\frac {1} {24} 
  E^c \wedge E^b \wedge E^a \epsilon_{abcd} \,\left(\bar{\sigma}{}^{d\dot{\alpha}\alpha}
  [\cald_\alpha, \bar{\cald}_{\dot\alpha}]\calp-3 G^d\calp \right)
 \, . \qquad
\end{aligned}
\ee
The super three-forms $(\cala_3^I,\tilde\cala_{3I})$ are  obtained by plugging the prepotentials \eqref{realprepot1} into \eqref{super3form}, using the  composite prepotential
\be\label{WZprepot} \calp = q_I \calp^I- p^I \tilde{{\cal P}}_I= 
-2q_I\calm^{IJ}\Im\Sigma_J +2p^I\Im(\bar\calg_{IJ}\calm^{JK}\Sigma_K)\; . \qquad 
\ee
The gauge transformations \eqref{gauge3form} here translate into the gauge transformations 
\be
\calp^I\rightarrow \calp^I+2L^I\quad,\quad \tilde\calp_I\rightarrow \tilde\calp_I+2\tilde L_I \, , 
\ee
where we recall that $(L^I, \tilde L_J)$ are arbitrary real linear multiplets.
  For the three--forms $\cala_{3I}$ and $\tilde{\cala}_{3}^I$ these transformations of the prepotentials 
determine the special structure  of the super-2-form parameters 
$\alpha_{2I}$ and $\tilde{\alpha}_{2}^I$
of the superspace gauge transformations $\cala_{3I}\mapsto \cala_{3I}+\d\alpha_{2I}$ and $\tilde{\cala}_{3}^I\mapsto \tilde{\cala}_{3}^I+\d\tilde{\alpha}_{2}^I$. Thus the WZ term is gauge invariant modulo boundary terms which vanish in the case of the closed supermembrane (or for an infinite  domain wall type object). Note that 
\eqref{super3form} and hence the WZ term \eqref{susyWZ} are also Weyl invariant by construction (for the coupling of the membrane to pure three-form supergravity this fact was noticed in \cite{Kuzenko:2017vil}).

The prepotentials and the super three-forms organize in symplectic vectors
\be
\left(\begin{array}{c}
\calp^I \\ \tilde\calp_{J}
 \end{array}\right)\quad,\quad \left(\begin{array}{l}
\cala^I_3 \\ \tilde\cala_{3J}
 \end{array}\right)
\ee
linearly transforming  under \eqref{defcalS}. We see that the complete WZ term \eqref{susyWZ} is invariant under the symplectic transformations provided that the vector of the charges
\be\label{chargev}
\left(\begin{array}{c}p^I \\ q_J
 \end{array}\right)
\ee
transforms as a symplectic vector as well. 
 Thus symplectic transformations cannot be considered as a symmetry of a single membrane characterized by definite values of the charges  $p^I$ and $q_I$, but rather of the whole set of supermembranes with all possible values of charges.  
A quantization of the membrane charges, which is automatic in stringy models, imposes the symplectic transformations \eqref{defcalS} to take discrete values. For instance, if $p^I,q_J\in\mathbb{Z}$, then $\cals\in$ Sp$(2n+2;\mathbb{Z})$ 
relate supermembranes with different allowed values of integer charges.

Now, the WZ term \eqref{susyWZ} should be completed with a supersymmetric NG-like term. Furthermore, we require the complete membrane action to be invariant under local $\kappa$-symmetry, in order to get a  supersymmetric physical  spectrum on the membrane world-volume. The appropriate NG-term turns out to be
\be\label{susyNG}
S_{\rm NG}=-2\int_\calc\d^3\xi\sqrt{-\det h}\left|q_IS^I-p^I\calg_I(S)\right| . 
\ee
In \eqref{susyNG} the bulk superfields $S^I$ are evaluated on $z^M(\xi)$ and $\det h\equiv \det h_{ij}(\xi)$ where
\be
h_{ij}(\xi)\equiv \eta_{ab}E^a_i(\xi)E^b_j(\xi)\quad~~\text{with }\quad E^a_i(\xi)\equiv E^a_M(z(\xi))\del_i z^M(\xi) . 
\ee

By using the standard constraints for the supervielbeins $E^A_M$, one can check that the complete action
\be\label{complmembrane}
S_{\rm M}\equiv S_{\rm NG}+S_{\rm WZ}
\ee
is invariant under the $\kappa$-symmetry transformations
\be\label{kappasymm}
\delta z^M (\xi)=\kappa^\alpha(\xi) E^M_\alpha(z(\xi))+\bar\kappa^{\dot\alpha}(\xi) E^M_{\dot\alpha}(z(\xi)). 
\ee
The local fermionic parameter $\kappa^\alpha(\xi)$ (with $\bar\kappa^{\dot\alpha}(\xi)\equiv \overline{\kappa^\alpha}(\xi)$) satisfies the projection condition 
\be\label{kappaproj}
\kappa_\alpha=\frac {q_IS^I-p^I\calg_I}{|q_IS^I-p^I\calg_I|}\Gamma_{\alpha\dot\alpha}\bar\kappa^{\dot\alpha} , 
\ee
where 
\be\label{kappagamma}
\Gamma_{\alpha\dot\alpha}\equiv \frac{\ii\epsilon^{ijk}}{3!\sqrt{-\det h}}\epsilon_{abcd} E^b_iE^c_j E^d_k\,\sigma^a_{\alpha\dot\alpha}. 
\ee

The proof of the invariance of the action under \eqref{kappasymm} is given in Appendix \ref{app:kappa}. The $\kappa$-symmetry implies that half of the degrees of freedom of the fermionic world-volume fields $\theta^\alpha(\xi)$ are pure gauge. On the other hand, the invariance under world-volume diffeomorphisms implies that three degrees of freedom contained in the bosonic fields $x^m(\xi)$ are pure gauge. Hence, the membrane carries one bosonic and two real fermionic degrees of freedom, which constitute the spectrum of an $\caln=1$ scalar supermultiplet in three dimensions.   

Actually, 
for the membrane action to be kappa-symmetric, it is sufficient to require that $\calg_I(S)\equiv\calg_{IJ}(S)S^J$ with no other 
restrictions on $\calg_{IJ}(S)$ (e.g. homogeneity restrictions). In other words, supermembranes can couple to more general classes of supergravity models than those we have concentrated our attention on. Other generalizations are described in Appendix \ref{app:gener}.

Note also that the bosonic contribution of the NG term \eqref{susyNG} exactly reproduces the field-dependent tension \eqref{tension} expected from string compactifications. Clearly, the supermembrane action \eqref{complmembrane} does not change under the (discrete) symplectic transformations, provided that \eqref{chargev} transform as a symplectic vector.  So, as we have already discussed, the  symplectic transformations should be considered as dualities  relating supermembranes with different (quantized) charges 
$(q_I, p^J)$.

Furthermore, as expected, the projector \eqref{kappaproj}  is compatible with the  projector associated with BPS membranes obtained by wrapping  probe D-branes in $\caln=1$ string compactifications, see for instance \cite{Martucci:2005ht,Martucci:2006ij} \footnote{The $\kappa$-symmetry of a $p$-brane was shown to be in one-to-one correspondence with supersymmetry  preserved by the $p$-brane BPS state \cite{Bergshoeff:1997kr} as well as by the `complete but gauge fixed' Lagrangian description of the supergravity coupled to $p$-brane interacting systems \cite{Bandos:2001jx,Bandos:2002kk,Bandos:2005ww}. }. However, we stress that the invariance under \eqref{kappasymm} goes beyond the probe regime, since it does not require the bulk sector to be on-shell. In other words, by summing \eqref{complmembrane} and \eqref{3formlagr} one gets a supersymmetric action describing the {\em off-shell}  coupling between the supergravity bulk and the membranes \footnote{Previous examples of  four-dimensional superfield actions of this type have been constructed  for  dynamical interacting systems of supergravity and/or matter multiplets coupled to  massless superparticles 
\cite{Bandos:2002bx}, superstrings \cite{Bandos:2003zk} and  supermembranes \cite{Bandos:2011fw,Bandos:2010yy}.}. 

Finally, so far the membrane action has been written in the Weyl-invariant and manifestly supersymmetric form. We can now gauge fix the Weyl invariance as described in Section \ref{sec:subclass}, write down the action in components and perform the standard Weyl rescaling $e^a_m\rightarrow e^{\frac16\calk}e^a_m$ for passing to the Einstein frame. By isolating the bosonic terms, one easily gets 
\be\label{compmembrane}
\begin{aligned}
S_{\rm M}=&\,-2\int_\calc \d^3\xi\sqrt{-\det h}\,\; e^{\frac12\calk}\left| q_If^I(\phi)-p^I\calg_I (\phi)\right|\\
&+q_I\int_\calc A_3^I-p^I\int_\calc\tilde A_{3I}+\text{(fermions)} \, , 
\end{aligned}
\ee
\sloppy{where now $h_{ij}$ denotes the pull-back of the bulk Einstein-frame metric, $h_{ij} \equiv {g_{mn}\del_i x^m\del_j x^n}$.}

In what follows we will gauge fix worldvolume reparametrization invariance of the action (\ref{compmembrane}) by imposing the static gauge, in which the worldvolume coordinates are identified with three of four coordinate functions $x^m(\xi)=(x^\mu(\xi), y(\xi))$, 
namely 
\be\label{staticG}
x^\mu(\xi)=\delta_i^\mu\xi^i\; , \qquad \mu=0,1,2\; . 
\ee 
In this gauge the worldvolume dynamics of the membrane is described by a single real field $y(x)$ which  
is a Goldstone field associated with the bulk diffeomorphism symmetry spontaneously broken by the membrane.


\section{Jumping domain walls}
\label{sec:jumpingDW}

In this section we study $\frac12$-supersymmetric solutions of a bulk-plus-membrane system. We will focus  on the class of models described in Section \ref{sec:subclass}. The extension to more general matter-coupled models is straightforward \footnote{ See \cite{Ovrut:1997ur} for an example of a supergravity domain wall in the presence of a single gauge three-form and no flowing scalar fields. 
Domain wall solutions in 4D minimal supergravity were discussed in \cite{Cvetic:1992bf,Cvetic:1996vr} and in \cite{Huebscher:2009bp}, where the duality equations relating scalars to 3-forms were  imposed.}.

Before concentrating on supersymmetric domain walls, let us analyze the bosonic sector of the theory depending on the three-forms. It includes the last two terms in \eqref{boscomp} and the  WZ term in \eqref{compmembrane}. 
Let us rewrite these terms as follows 
\be\label{bm3forms}
\begin{aligned}
&\int \d^4x\,e\,\calt^{IJ}{}^*\!\bar\calf_{4I}\, {}^*\!\calf_{4J}-\int (p^I \tilde A_{3I}-q_I A_3^I)\wedge\delta_1(\calc)\\ &~~~~~+2\int_\calb\tilde A_{3I}\,\Re(\calt^{IJ}{}^*\!\calf_{4J})-2\int_\calb A^I_{3}\,\Re(\calg_{IJ}\calt^{JK}{}^*\!\calf_{4K}) \, , 
\end{aligned}
\ee
where $\delta_1(\calc)$ is a delta-like one-form localized on the membrane world-volume $\calc$. In the static gauge (\ref{staticG}) 
 $\delta_1(\calc) = \d y\delta (y-y(x))$ and in the bulk diffeomorphism  gauge (\ref{y=0}) it reduces to   $\delta_1(\calc)= \d y\delta (y)$.

Varying \eqref{bm3forms} we get the three-form equations of motion 
\be\label{fluxeq2}
\begin{aligned}
\d\Re\left(\calt^{IJ}{}^*\!\calf_{4J}\right)&=-\frac12p^I\delta_1(\calc),\\
\d\Re\left({\calg}_{IJ}\calt^{JK}{}^*\!\calf_{4K}\right)&=-\frac12q_I\delta_1(\calc).
\end{aligned}
\ee
Comparing them with \eqref{fluxeq}, we see that the membrane has modified the latter by localized sources   proportional to the charges $(q_I,p^J)$.  This means that the solution of \eqref{fluxeq2} is still given by \eqref{incost}, with constant $(m^I,e_J)$ away from the membrane. On the other hand, as one passes from  the left to the right of the membrane, with respect to the orientation defined by the one-form $\delta_1(\calc)$, the values of these constants `jump' as follows
\be\label{fluxjump}
m^I\rightarrow m^I-p^I\,,\quad e_I\rightarrow e_I-q_I.
\ee
This implies that we may still integrate out the three-forms away from the membrane, getting an ordinary supergravity with the superpotential \eqref{Esup}. However, we should at least use two such ordinary supergravity actions, one on the left and one on the right from the membrane worldvolume, whose superpotentials are related by the jump  \eqref{fluxjump}.

If the  three-form gauge symmetries are compact  as discussed in Section \ref{sec:quant}, from \eqref{fluxjump} and \eqref{bulkquant}  we immediately conclude that the membrane charges must be integrally quantized
\be
p^I, q_J\in\mathbb{Z}\,.
\ee

Let us now come back to the  search for flat domain walls  including the membrane.  We split  the space-time coordinates $x^m$ into $(x^\mu,x^3\equiv y)$, $\mu=0,1,2$, and take the following ansatz for the space-time metric 
\be\label{ds2=ansatz}
\d s^2=e^{2D(y)}\d x^\mu\d x_\mu+\d y^2 \, . 
\ee
We would like to study the simplest supersymmetric domain wall associated with a single flat membrane located at $y=0$, that is $\calc=\{y=0\}$. The extension  to the case of multiple membranes is straightforward.

The fermions are set to zero and the scalar fields $\phi^i$ are allowed to depend only on the transverse coordinate 
\be\label{scansatz}
\phi^i=\phi^i(y).
\ee
As we will see, it is consistent to assume that $ \phi^i(y)$ are continuous in $y$, while their derivative may jump at $y=0$.

In the class of models described in Section \ref{sec:subclass}, the presence of the chiral multiplets $T^p$ would  not allow for supersymmetric vacua with $W\neq 0$.  This would imply that a BPS domain wall must necessarily degenerate on one or both of its sides  (see for instance \cite{Cvetic:1994px}). 
Hence, in order to lighten the  discussion,  we will assume absence of $T^p$ multiplets, which may be easily reinstated into the flow equations. We can then write the complete K\"ahler potential in the form 
\be\label{compKb}
\calk(\Phi,\bar\Phi)=K(\Phi,\bar\Phi)+\hat K_0 \, , 
\ee
where $\hat K_0$ is a constant.    

For the three-form gauge potentials one chooses an ansatz  which respects the symmetries of the domain wall configuration setting
\be\label{3form=DW}
A^I_3=\alpha^I(y)\d x^0\wedge \d x^1\wedge \d x^2\,,\quad \tilde A_{3I}=\tilde\alpha_I(y)\d x^0\wedge \d x^1\wedge \d x^2,
\ee
which are assumed to be continuous at the membrane position $y=0$.

\subsection{Bulk supersymmetry}
\label{sec:bulksusy}

To find the flow equations satisfied by $\frac12$-BPS domain walls,  one imposes that the corresponding Killing equations admit two independent solutions. As usual, the Killing equations are obtained by imposing that the supersymmetry transformations of the fermions vanish, and can be found in Appendix \ref{app:susy} (see equation \eqref{susydw}). Their analysis  is carried out in a way similar to the derivation of the flow equations for the domain walls in standard supergravity (see e.g. \cite{Cvetic:1992bf, Cvetic:1992st, Cvetic:1992sf, Cvetic:1993xe,Cvetic:1996vr,  Ceresole:2006iq} for details). As a result one gets the following flow equations 
\begin{subequations}\label{floweq1}
\begin{align}
\dot \phi^i &= e^{\frac 12\calk(\phi,\bar\phi)+i\vartheta(y)}K^{i\bar\jmath}(\overline W_{\bar\jmath}+K_{\bar\jmath} \overline W)\,,\label{susy22}\\
\dot D &= - e^{\frac 12\calk(\phi,\bar\phi)}|W|,\label{susy11}
  \end{align}
\end{subequations}
where the dot corresponds to the derivative with respect to $y$, e.g.\ $\dot D \equiv \frac{\d }{\d y}D$, and   $\vartheta(y)= \vartheta(\phi(y),\bar{\phi}(y))$ is the phase of $W$, 
namely 
\be\label{vartheta} 
W=e^{i\vartheta}|W|\,.
\ee 
We recall that, before fixing the expectation value of the field-strengths, $W$ and $W_i$ are defined as in \eqref{W=}.

Note that the supersymmetry  
 preserved by the domain wall is characterized by
the Killing spinor $\zeta_\alpha(y)$ satisfying the projection condition
\be\label{spinproj}
\zeta_\alpha= \ii e^{\ii\vartheta}(\sigma_{\ul 3})_{\alpha\dot\alpha}\bar\zeta^{\dot\alpha}.
\ee
In order to have an everywhere supersymmetric domain wall solution, including the membrane sitting at $y=0$, it is natural to require $\vartheta(y)$ to be continuous at $y=0$. From the first equation in \eqref{susydw} and \eqref{spinproj} it then follows that
\be\label{thetaflow}
 \dot\vartheta=-\Im\left(\dot\phi^i K_i\right) .
\ee

Now, for the complete bulk description of the domain wall configurations we should add to \eqref{floweq1} the equations of motion of the three-forms sourced by the membrane \eqref{fluxeq2}. For the domain wall configuration their solution gives the following form of the superpotential $W$ 
\be\label{Wmem}
W(\phi,y)=  e_{-I}f^I(\phi)-m_-^{I}\calg_I(\phi)-\Theta(y)\left(q_If^I(\phi)-p^I\calg_I(\phi)\right) \, , 
\ee
where $ \Theta(y)$ is the Heaviside step function, 
and 
\be 
W_i=\partial_iW . 
\ee 
Comparing this form of $W$ with Eq.\,\eqref{Esup} in the absence of the membrane we see that the superpotential becomes a step function of $y$. 
This means that away from the membrane at $y=0$  the bulk fields obey the standard supergravity equations associated with two distinct superpotentials $W_-$ and $W_+$, for $y<0$ and $y>0$ respectively. These superpotentials take the form  \eqref{Esup} associated with the constants $(m^I_-,e_{-I})$ and $(m^I_+,e_{+I})$, satisfying the relations $m^{I}_-=m^{I}_++p^I$ and $e_{-I}=e_{+I}+q_I$. 

Let us now introduce a `flowing' covariantly holomorphic superpotential \cite{Ceresole:2006iq} 
\be\label{jumpc}
\calz(\phi,y)\equiv e^{\frac12 \calk(\phi,\bar\phi)}W=e^{\frac12 \calk(\phi,\bar\phi)}\left[\Theta(y)W_+(\phi)+\Theta(-y)W_-(\phi)\right].
\ee
As in \eqref{Wmem}, the dependence of $\calz$ on $y$ is both explicit, through the step functions, and implicit,  through \eqref{scansatz}. The membrane induces the jump 
\be\label{calzjump}
\Delta\calz \equiv\lim_{\varepsilon\rightarrow  0}\left(\calz|_{y=\varepsilon}-\calz|_{y=-\varepsilon}\right)= -e^{\frac12 \calk}\left(q_If^I-p^I\calg_I\right)|_{y=0}.
\ee
The absolute value of $\Delta \calz$ is determined by the membrane tension
\be\label{effTM}
T_{\rm M}\equiv 2\,e^{\frac12 \calk}\left|q_If^I-p^I\calg_I\right|_{y=0}=2|\Delta\calz| 
\ee
and at $y=0$ the phase $e^{i\vartheta(y)}$  enters the $\kappa$-symmetry projector \eqref{kappaproj}. 

Due to the holomorphicity of $W$, which implies 
\be\label{partialmod}
 \partial_{\bar{\jmath}}|W|+i\partial_{\bar{\jmath}}\vartheta \; |W|=0,
\ee
and the definition of $\calz$, we have
\be\label{dZ}
\del_{\bar\jmath}|\calz|\equiv \frac12 e^{\ii\vartheta}e^{\frac12 \calk}\bar D_{\bar\jmath}\bar W\,.
\ee
Hence, in terms of $\calz$ the flow equations \eqref{floweq1} take a known form
 \cite{Cvetic:1992bf, Cvetic:1992st, Cvetic:1992sf, Cvetic:1993xe,Cvetic:1996vr,  Ceresole:2006iq} 
\begin{subequations}\label{floweq}
\begin{align}
\dot \phi^i &= 2K^{i\bar\jmath}\,\del_{\bar\jmath}|\calz|,\label{susy2}\\
\dot D &= - |\calz|,\label{susy1}
  \end{align}
\end{subequations}
where $K^{i\bar\jmath}$ is the inverse of the $\phi^i$ kinetic matrix which in our case is $K_{i\bar\jmath}\equiv f_i^I\bar f^J_{\bar\jmath}G_{IJ}$, with $G_{IJ}$ defined in (\ref{GIJ}). 

We see that, owing to \eqref{dZ}, the flow equation \eqref{susy2} has fixed-point solutions provided by the supersymmetric vevs $\phi^i_*$ (such that $D_{\bar\jmath}W|_{\phi_*}=0$).  Then the solution of Eq.\,\eqref{susy1}  is $D=-|\calz_*|y+\text{const.}$, which corresponds to an AdS space of radius $1/|\calz_*|$ for $\calz_*\neq 0$ and to flat space if $\calz_*=0$. Hence, a  regular  BPS domain wall  interpolates between two different supersymmetric vacua and its geometry is asymptotically AdS or flat for $y\rightarrow\pm\infty$.      

As it will be clear from the following discussion, for the given choice of signs, the flow equations \eqref{floweq} lead to a complete 
solution only if $\calz$ is nowhere vanishing along the flow and if $|\calz|_{y=+\infty}\not= |\calz|_{y=-\infty}$. We will assume    $|\calz|_{y=+\infty}>|\calz|_{y=-\infty}$,  but, by  the coordinate  flip $y\rightarrow -y$,  one can analogously consider the case $|\calz|_{y=+\infty}<|\calz|_{y=-\infty}$ (and nowhere vanishing $\calz$). The latter case then requires opposite signs in \eqref{floweq} and other flow equations. Note also that  performing the change $y\rightarrow -y$, one should also flip the relative sign of the Killing spinor projector \eqref{spinproj}. This should be correlated with the sign in the corresponding kappa-symmetry projector which, in turn, is related to the sign of the membrane WZ term (see Section \ref{sec:+membrane}).

As we have already mentioned, the phase $\vartheta(y)$ is required to be a continuous function.
In other words, we require the phase of $\calz$ not to change in passing through the domain wall, so that $\Delta \calz=e^{\ii\vartheta(0)}|\Delta \calz|$. 
This requirement together with the covariant holomorphicity of $\calz$ and Eq.\,\eqref{susy2}  imply that  $\vartheta(y)$ satisfies the flow equations \eqref{thetaflow} and hence the following identity holds
\be\label{diffcalz}
\frac{\d |\calz|}{\d y}=2\Re\left(\dot\phi^i\del_i|\calz|\right) +\frac12 T_{\rm M}\,\delta(y),
\ee
which just follows from \eqref{jumpc}. 
Using the flow equation \eqref{floweq1}, we can rewrite \eqref{diffcalz}  in the following form
\be\label{extZ}
 2 e^{-3D} \frac{\d }{\d y}\left(e^{3D}|\calz|\right)=-e_I(y) {}^*\!F_4^I+m^I(y){}^*\!\tilde F_{4I}- T_{\rm M}\delta(y),
\ee
where 
\be\nonumber
m^I(y)\equiv m^I_--p^I\Theta(y) \ , \qquad e_I(y)\equiv e_{-I}-q_I\Theta(y).
\ee 
Equation \eqref{extZ} is solved by choosing (continuous) gauge potentials $A^I_3,\tilde A_{3J}$ such that 
\be
2|\calz|{\rm vol}_3=e_I(y) A^I_3-m^I(y)\tilde A_{3I}\,,
\ee
with ${\rm vol}_3\equiv e^{3D}\d x^0\wedge \d x^1\wedge \d x^2$,
which in turn implies that
\be\label{T=A}
T_{\rm M}{\rm vol}_3|_{y=0} =\left(q_I A^I_3-p^I\tilde A_{3I}\right)|_{y=0}\,.
\ee
Hence, there is a perfect cancellation between the on-shell values of the NG and WZ term in the membrane action evaluated on the domain wall solution. This is somewhat similar to static membrane solutions in $AdS\times S$ backgrounds \cite{Claus:1998mw}.

As in \cite{Ceresole:2006iq}, we can combine \eqref{floweq} and \eqref{diffcalz} to get the following equation  
\be\label{Ctheorem}
\dot C=4K^{i\bar\jmath}\del_i|\calz|\del_{\bar\jmath}|\calz|+\frac12T_{\rm M}\delta(y)\geq 0 , 
\ee
where we have introduced $C(y)\equiv -\dot D(y)$. 
The function $C(y)$ is analogous to the monotonically increasing c-function introduced in \cite{Girardello:1998pd,Freedman:1999gp} in the AdS/CFT context. Equation \eqref{Ctheorem} shows the contribution of the membrane to the monotonic flow of $C(y)$, which `jumps up' by $\frac12T_M$ at $y=0$. A similar equation was also derived in \cite{Haack:2009jg} by dimensionally reducing ten-dimensional flow equations in the presence of effective membranes corresponding to D-branes wrapped along internal cycles.    

Under our assumption that $|\calz|_{y=+\infty}>|\calz|_{y=-\infty}$, equation \eqref{susy1} tells us that $|\calz|=C$. Hence \eqref{Ctheorem} also implies that $|\calz|$ monotonically increases as we move from $y=-\infty$ to $y=+\infty$. Clearly, in the case $|\calz|_{y=+\infty}<|\calz|_{y=-\infty}$, the appropriate sign-reversed flow equations imply that $|\calz|$ is monotonically decreasing, while \eqref{Ctheorem} still holds, since in that case the sign-reversed \eqref{susy1} becomes $|\calz|=\dot D\equiv -C$.

Finally, let us momentarily remove our assumption that $\calz$ is nowhere vanishing, supposing that $\calz|_{y_0}=0$ at some transversal coordinate $y_0$. Then, $|\calz|$ must necessarily be monotonically increasing for $y>y_0$ and decreasing for $y<y_0$, so that $\calz$ can vanish only at $y_0$. This implies that for $y>y_0$ the flow equations \eqref{floweq} hold, while for $y<y_0$ one must use the sign-reversed ones.   

In conclusion, the bulk supersymmetry conditions \eqref{spinproj}, \eqref{thetaflow}, \eqref{susy2} and \eqref{susy1}  are basically unaffected by the presence of the membrane. In the following subsection we will re-derive them from an effective one-dimensional BPS action.  


\subsection{BPS action and domain wall tension}

The above supersymmetric flow equations can be alternatively derived by plugging the above domain wall ansatz into the complete bulk-plus-membrane action and using the three-form equations of motion.

Let us first focus on the terms appearing in \eqref{bm3forms}. Using  Stokes' theorem, we can rewrite them as follows, 
\be\label{bm3forms2}
\begin{aligned}
&-\int \d^4x\,e\,\calt^{IJ}{}^*\!\bar\calf_{4I}\, {}^*\!\calf_{4J}+\int  A^I_{3}\wedge \left[2\d\Re( {\calg}_{IJ}\calt^{JK}{}^*\!\calf_{4K})+q_I \delta_1(\calc)\right]\\
&~~~~~~~~~~~~~-\int \tilde A_{3I}\wedge \left[2\d\Re(\calt^{IJ}{}^*\!\calf_{4J})+p^I \delta_1(\calc)\right].
\end{aligned}
\ee
It is then easy to see that, if we integrate out the gauge three-forms using their equations of motion \eqref{fluxeq2}, we are left with the following term 
\be\label{potential2}
 -\int \d^4x\,e\,\calt^{IJ}{}^*\!\bar\calf_{4I}\, {}^*\!\calf_{4J} \quad~~~\text{(on-shell 3-forms)} . 
\ee
We can then repeat the discussion of Section \ref{sec:equiv}, writing \eqref{potential} as the (minus) potential of the standard $\caln=1$ supergravity (see Eq.\,\eqref{potst}), with the only difference that the superpotential is not constant but changes as described above when passing the membrane position $y=0$. Then, for the domain wall solution under consideration,  Eq.\,\eqref{potential2} can be written in terms of the jumping central charge \eqref{jumpc} as follows  
\be\label{potZ}
\int \d^4x\,e\,\calt^{IJ}{}^*\!\bar\calf_{4I}\, {}^*\!\calf_{4J}=\int\d^3 x\int\d y\, e^{3D} \left(K^{i\bar\jmath}D_i\calz\bar D_{\bar\jmath}\bar\calz-3|\calz|^2\right),
\ee
where $D_i\calz\equiv \del_i\calz+\frac12 K_i \calz$. 

Let us now consider the remaining,  `gravitational' part of the bosonic action, which is given by
\be\label{gravaction}
\begin{aligned}
&-\int \d^4x\,e\, \Big(\frac{1}{2}R+ K_{i\bar\jmath}\,  \del \phi^i \del\bar{\phi}^{\bar \jmath}\Big) +S_{\rm GH}\\
&~~~~~~~~~ -2\int_\calc \d^3\xi\sqrt{-\det h}\,\; e^{\frac12 \calk}\left| q_If^I(\phi)-p^I\calg_J(\phi)\right|,
\end{aligned}
\ee
where $S_{\rm GH}$ is the Gibbons--Hawking boundary term \cite{Gibbons:1976ue}. 

For the domain wall ansatz 
in the static gauge \eqref{staticG}, the contribution of the membrane at $y=0$ appearing in the second line of \eqref{gravaction} reduces to 
\be\label{gravmem}
-\int\d^3 x\int \d y\,  \delta(y)T_{\rm M}e^{3D},
\ee
where $T_M$ has been defined in \eqref{effTM}.

Now, following \cite{Ceresole:2006iq}, one can take the sum of \eqref{potZ} with the first line of \eqref{gravaction} evaluated on the domain wall ansatz and write it in the form
\be\label{aBPSgrav}
\begin{aligned}
&\int\d^3 x\int\d y\, e^{3D}\left[3\big(\dot D+|\calz|\big)^2-K_{i\bar\jmath}\big(\dot\phi^i -2K^{i\bar k}\del_{\bar k}|\calz|\big)\big(\dot{\bar\phi}^{\bar\jmath} -2K^{l\bar \jmath}\del_l|\calz|\big)\right]\\
&~~~~~~-2\int\d^3 x\int\d y\, e^{3D}\left[3\dot D|\calz|+2\Re\big(\dot\phi^i\del_i|\calz|\big)\right].
\end{aligned}
\ee
On the other hand, using \eqref{diffcalz} we can write the second line of \eqref{aBPSgrav} in the form
\be
-2\int\d^3 x\int\d y\,\left[ \frac{\d}{\d y}\big(e^{3D}|\calz|\big) -\frac12 \delta(y)T_{\rm M}e^{3D}\right]
\ee
whose second term is precisely the opposite of \eqref{gravmem}. 

Then, in the sum of \eqref{aBPSgrav} and \eqref{gravmem}, the terms localised on the membrane perfectly cancel  and the complete action reduces to the following BPS form
\be\label{redS}
\begin{aligned}
S_{\rm red}=&\,\int\d^3 x\int\d y\, e^{3D}\left[3\big(\dot D +|\calz|\big)^2-K_{i\bar\jmath}\big(\dot\phi^i-2K^{i\bar k}\del_{\bar k}|\calz|\big)\big(\dot{\bar\phi}^{\bar\jmath}-2K^{l\bar \jmath}\del_l|\calz|\big)\right]\\
&-2\int\d^3 x\left[\big( e^{3D}|\calz|)|_{y=+\infty}-\big(e^{3D}|\calz|\big)|_{y=-\infty}\right].
\end{aligned}
\ee
This reduced action is identical in form to the one obtained in \cite{Ceresole:2006iq} in the absence of membranes, basically because of the observed reciprocal cancellations of various terms localised on the membrane. 

Hence, as in the absence of membranes,  the extremization of the BPS action \eqref{redS} precisely reproduces the bulk flow equations \eqref{floweq}. 
Furthermore, on any solution of the flow equations, we get
\be\label{onshellred}
S_{\rm red}|_{\rm on-shell}=-2\int\d^3 x\left[\big( e^{3D}|\calz|)|_{y=+\infty}-\big(e^{3D}|\calz|\big)|_{y=-\infty}\right]=-\int\d^3 \tilde x \,T_{\rm DW} , 
\ee
where on the slices of constant $y$ we have introduced coordinates $\tilde x^\mu=e^{D(y)}x^\mu$, so that $\d^3 \tilde x$ denotes the physical volume, and  
\be\label{DWtension}
T_{\rm DW}=2\big(|\calz|_{y=+\infty}-|\calz|_{y=-\infty}\big)
\ee
denotes the overall tension of the domain wall.

Equation \eqref{DWtension} is formally identical to the formula obtained in the absence of membranes \cite{Cvetic:1992bf, Cvetic:1992st, Cvetic:1992sf, Cvetic:1993xe,Cvetic:1996vr,  Ceresole:2006iq}. However one should keep in mind that it includes the contribution of the membrane. 
This can be seen by splitting the overall change of $|\calz|$ in the bulk and membrane contributions 
\be\label{DWtension2}
T_{\rm DW}=2\big(|\calz|_{y=+\infty}-\lim_{\varepsilon\rightarrow 0}|\calz|_{y=\varepsilon}\big)+2\big(\lim_{\varepsilon\rightarrow 0}|\calz|_{y=-\varepsilon}-|\calz|_{y=-\infty}\big)+ T_{\rm M}\,.
\ee
See also \cite{Haack:2009jg} for the same conclusion reached starting from a  ten-dimensional description of similar domain wall solutions. 

From \eqref{DWtension} we see that our working assumption $|\calz|_{y=+\infty}> |\calz|_{y=-\infty}$ guarantees that $T_{\rm DW}>0$. The case  $|\calz|_{y=+\infty}< |\calz|_{y=-\infty}$ (with still nowhere vanishing $\calz$)  can be obtained by changing $y\rightarrow -y$ in the above steps, so that the sign-reversed flow equations \eqref{floweq} extremize the corresponding BPS reduced action and the tension is given by $T_{\rm DW}=2\big(|\calz|_{y=-\infty}-|\calz|_{y=+\infty}\big)$. 
 
Furthermore, as mentioned at the end of Section \ref{sec:bulksusy}, the case in which there is a vanishing point of $y_0$ of $\calz$ can be obtained by gluing two regions along which $|\calz|$ flows in opposite directions, first decreasing from $|\calz|_{y=-\infty}$ to $0$ and then increasing to $|\calz|_{y=+\infty}$. The above arguments can be easily adapted to this case as well and give  $T_{\rm DW}=2\big(|\calz|_{y=-\infty}+|\calz|_{y=+\infty}\big)$, again as in the absence of membranes \cite{Cvetic:1992bf, Cvetic:1992st, Cvetic:1992sf, Cvetic:1993xe,Cvetic:1996vr, Ceresole:2006iq}. However,  in this case the membrane sitting at $y_0$ would have vanishing localized tension, $T_M=0$. This would signal breaking of the validity of the effective action.   


\subsection{World-volume analysis}

It remains to check that the membrane world-volume preserves the same supersymmetry as the domain wall solution of the  bulk field equations, and that the membrane equations of motion are satisfied.

As discussed above, we can consider $\calz$ to be nowhere vanishing and  $|\calz|_{y=+\infty}> |\calz|_{y=-\infty}$, without loss of generality.  Recall that the Killing spinors $\zeta_\alpha(y)$ satisfy the projection condition  \eqref{spinproj}. These bulk supersymmetries  act on the membrane sector by shifting the world-volume fermions.  Hence, they are preserved by the membrane only if they can be regarded as (gauge) $\kappa$-transformations. We should hence check that on the membrane 
\be\label{zetakappa}
\zeta_\alpha|_{y=0}=\kappa_\alpha
\ee
with $\kappa_\alpha$ satisfying the constraint \eqref{kappaproj}. Due to the global continuity of the phase of  $\calz$, in the case at hand the condition \eqref{kappaproj} can be written as
\be\label{kappaproj2}
\kappa_\alpha=\frac{\Delta \calz}{|\Delta\calz|}\Gamma_{\alpha\dot\alpha}\bar\kappa^{\dot\alpha}=e^{\ii\vartheta}\Gamma_{\alpha\dot\alpha}\bar\kappa^{\dot\alpha},
\ee
where  $\Gamma_{\alpha\dot\alpha}$ is defined in \eqref{kappagamma} and should be evaluated for the static membrane placed at $y=0$.  This gives 
\be\label{sigma3}
\Gamma_{\alpha\dot\alpha}=\ii (\sigma_{\ul 3})_{\alpha\dot\alpha} \, . 
\ee
We see that
\eqref{kappaproj2} with \eqref{sigma3} is equivalent to the restriction of
\eqref{spinproj} to $y=0$. This implies that the membrane world-volume is perfectly compatible with the background supersymmetry. Hence the fully coupled bulk-plus-membrane domain wall configuration preserves two supersymmetries out of four. 

 From the form of the BPS action of the previous subsection, we have seen that any explicit dependence on the membrane has disappeared. Moreover, in view of Eq.\,\eqref{T=A}, the on-shell membrane action vanishes. In other words, even if we move the position of the domain-wall membrane configuration,  the action \eqref{redS} and its on-shell value \eqref{onshellred} remain unchanged (for fixed boundary conditions). This clearly suggests  that the membrane equations of motion are identically satisfied for the considered domain wall solutions thus confirming their consistency. \footnote{This is in agreement with a general statement \cite{Gurses:1974cm,Aragone:1975ur,Bandos:2003tm} that the p-brane equations of motion in the interacting system including dynamical gravity can be obtained as a consistency condition for the Einstein  equations, i.e. the covariant energy-momentum tensor conservation.   
 }
 
 Indeed, in the static gauge and on the domain wall solution the membrane equations \eqref{mem-eq=Gen} reduce to
 \be\label{memeq}
 \left[e^{-3D}\frac\d{\d y}\left(e^{3D}|T|\right) - q_I{}^*\!F^{4I}+p^I{}^*\!\tilde{F}_{4I}\right]_{y=0}=0,
\ee
where $T(y)\equiv q_If^I(\phi(y))-p^I\calg_I(\phi(y))\equiv e^{i\vartheta_T(y)}|T(y)|$, so that $|T|_{y=0}=T_M$ and $\vartheta_T|_{y=0}$ coincides with the phase $\vartheta|_{y=0}$ of $\calz$ at $y=0$. Now, using the flow equations \eqref{floweq1}, upon some algebra one can check that the following relation holds for an arbitrary $y$
\be\label{bulkid}
e^{-3D}\frac\d{\d y}\left(e^{3D}|T|\right)-\cos(\vartheta-\vartheta_T)\left(q_I{}^*\!F^{4I}-p^I{}^*\!\tilde{F}_{4I}\right)-\sin (\vartheta-\vartheta_T) X=0,
\ee
with
\be
X= \calm^{IJ} \left[(q_I + \caln_{IL}p^L) {}^*\!\tilde{F}_{4J} + \left(\calm_{JK}\calm_{IL}p^L-\caln_{JK}\caln_{IL} p^L-\caln_{JK}q_I\right){}^*\!F_{4}^K\right]
\ee
and $\caln_{IJ}\equiv \Re \calg_{IJ}$. We see that even though the individual terms in \eqref{bulkid} are discontinuous at $y=0$, their combination is such that the limit $y\to 0$ of the whole expression is well-defined and produces the membrane equation of motion.


\section{A simple model}
\label{sec:example}

To exemplify the above general discussion, we now consider a concrete simple model. It has two double three-form multiplets $S^I=(S^0,S^1)$, associated with the prepotential 
\be\label{expre}
\calg=-\ii S^0 S^1.
\ee
Since $\calg$ is quadratic, the $2\times 2$ matrix
\be
\calg_{IJ}=\ii\calm_{IJ}=-\ii\left(\begin{array}{cc} 0 & 1\\
1 & 0\end{array}\right)
\ee
is constant, and the constraint \eqref{Sdef},  defining the chiral superfields $S^I$ in terms of the complex linear superfields $\Sigma_I=(\Sigma_0,\Sigma_1)$,  becomes linear
\be
S^0= - \frac\ii2(\bar\cald^2-8\calr)\Im\Sigma_1\,,\quad S^1=-\frac\ii2(\bar\cald^2-8\calr)\Im\Sigma_0 \, . 
\ee

Consider a general Lagrangian of the form \eqref{3formlagr}, putting to zero the spectators $T^p$ and the superpotential $\hat \calw$. At the component level, the Lagrangian includes four three-forms
\be\label{exgauge3}
\left(\begin{array}{l} A^0_{(3)} \\
A^1_{(3)}\\
\tilde A_{(3)0}\\
\tilde A_{(3)1}
\end{array}\right)
\ee
where, for clarity, we have introduced the change of notation $A^I_{3}\rightarrow A^I_{(3)}$, etc. As in \eqref{defcalf}, one can combine the corresponding field strengths $F^0_{(4)},F^1_{(4)},\tilde F_{(4)0}, \tilde F_{(4)1}$ into the complex field-strengths
\be\label{excalf}
\calf_{(4)0}=\tilde F_{(4)0}-\ii F_{(4)}^1\,,\quad \calf_{(4)1}=\tilde F_{(4)1}-\ii F_{(4)}^0 \, . 
\ee

Let us introduce the parametrization \eqref{SYPHI}  of $S$ in terms of the chiral superfields $(Y,\Phi)$, with 
\be\label{exfI}
f^0(\Phi)=1\,,\quad f^1(\Phi)=-\ii\Phi \,.
\ee
Then,  the general arguments  of \cite{Farakos:2017jme} imply that upon gauge-fixing the super-Weyl invariance and integrating out the gauge three-forms, one recovers standard supergravity coupled to the chiral superfield $\Phi$ with superpotential 
\be\label{exSup}
W(\Phi)=(e_0+\ii m^1)-\ii (e_1 +\ii m^0)\Phi \, ,
\ee
where $e_0,e_1,m^0,m^1$ are real constants. Notice that this result does not depend on the choice of the kinetic function $\Omega(S,\bar S)$ in the Lagrangian \eqref{3formlagr}. In the following we will focus on a specific simple choice for $\Omega(S,\bar S)$. 

\subsection{Bosonic action, three-forms  and $SL(2,\mathbb{Z})$ dualities}

Let us now take $\Omega(S,\bar S)$ as in \eqref{Omega0}, which corresponds to using for the scalar field $\phi=\Phi|$ the special K\"ahler potential
\be\label{exKP}
K(\phi,\bar\phi)=-\log\left(4\Im\phi\right)
\ee
associated with the prepotential $\calg$.
In particular, we see that in order to have a well defined K\"ahler potential we should require
\be\label{phipos}
\Im\phi>0.
\ee
Then, by the general results of Section \ref{sec:bosaction}, the bosonic sector of the supergravity action for $\phi$ and the gauge three-forms \eqref{exgauge3} is given by  
\be\label{exbaction}
S=-\int\d^4x\,e\,\left[\frac12 R+\frac{\del \phi\,\del\bar\phi}{4(\Im\phi)^2}-\calt^{IJ}(\phi){}^*\!\bar\calf_{4I}{}^ *\!\calf_{4J}\right]+S_{\rm bd}
\ee
with $S_{\rm bd}$ as in \eqref{bdaction} and 
\be\label{excalt}
\calt^{IJ}(\phi)= \frac{e^{-\hat K_0}}{6\,\Im\phi}\left(\begin{array}{rc} 1 {\ \ \ \ \ \ } & \ii\phi -  \Im\phi\\
-\ii\bar\phi -  \Im\phi & |\phi|^2\end{array}\right)
\, , 
\ee
where $\hat K_0$ is the constant appearing in the complete K\"ahler potential \eqref{compKb}.
It is known that the group of symmetries associated with the special K\"ahler structure defined by the prepotential \eqref{expre} is $SL(2,\mathbb{R})$ (see for instance \cite{Freedman:2012zz}). More precisely, an element
\be
\left(\begin{array}{rr} a & b\\
c & d\end{array}\right)\in SL(2,\mathbb{R}) 
\ee
acts on $\phi$ as follows
\be\label{phiSL2R}
\phi\rightarrow\frac{a\phi+b}{c\phi+d}\,.
\ee
In the assumption of the quantization conditions of Section \ref{sec:quant}, this reduces to $SL(2,\mathbb{Z})$ (with $a,b,c,d\in\mathbb{Z}$), which we will interpret as the duality group of the model. This duality symmetry $SL(2,\mathbb{Z})$ is embedded into the group of symplectic transformations $Sp(4,\mathbb{Z})$, see e.g. \cite{Freedman:2012zz}. Here we focus on the $SL(2,\mathbb{Z})$ generators 
\be
\mathfrak{t}=\left(\begin{array}{rr} 1 & 1\\
0 & 1\end{array}\right),\quad\mathfrak{s}=\left(\begin{array}{rr}\ 0 &\ 1\\
-1 & 0\end{array}\right) 
\ee
which correspond to the following $Sp(4,\mathbb{Z})$ transformations
\be\label{SL2ZS}
\cals(\mathfrak{t})=\left(\begin{array}{rrrr} 1 &\ 0 &\ 0 &\ 0\\
0 & 1 & 0 & 1\\
-1& 0 & 1 & 0\\
0 & 0 & 0 & 1
\end{array}\right),
\quad\cals(\mathfrak{s})=
\left(\begin{array}{rrrr} 
 0 &\ 0 &\ 1 &\ 0\\
0 & 0 & 0 & 1\\
-1& 0 & 0 & 0\\
0 &-1 & 0 & 0
\end{array}\right)\,.
\ee
One can check that,  applying $\cals(\mathfrak{t})$ and $\cals(\mathfrak{s})$ to the (bosonic component of)  the symplectic vector \eqref{symplV2}, with $f^I(\Phi)$ as in \eqref{exfI}, one gets the $\mathfrak{t}$- and $\mathfrak{s}$-actions on $\phi$ as in \eqref{phiSL2R}.\footnote{Under $\mathfrak{s}$ one needs to make also the change  $Y\rightarrow \phi Y$, which can be reabsorbed by a K\"ahler transformation $K\rightarrow K-\log\phi -\log\bar\phi$.} Furthermore, one can verify that $[\cals(\mathfrak{s})\cals(\mathfrak{t})]^3=\bbone$, which together with $\cals(\mathfrak{s})^2=-\bbone$  implies that \eqref{SL2ZS}  generate a four-dimensional representation of $SL(2,\mathbb{Z})$. 

By applying \eqref{SL2ZS} to \eqref{exgauge3}, one can then get the transformation properties of the gauge three-forms under the $SL(2,\mathbb{Z})$ duality group. Consider  the complex field-strengths \eqref{excalf} and organize them into a 2-components vector
\be\label{calfvec}
\vec\calf_{(4)}\equiv \left(\begin{array}{r}
\calf_{(4)0}\\
\calf_{(4)1}
\end{array}\right) \, . 
\ee
Under  $SL(2,\mathbb{Z})$ we have  $\vec\calf_{(4)}\rightarrow U\vec\calf_{(4)} $, with
\be\label{USL(2,Z)}
U(\mathfrak{t})= \left(\begin{array}{rr} 
1 & -\ii\\
0 & 1
\end{array}\right) 
\,,\quad 
U(\mathfrak{s})=
\left(\begin{array}{rr}
0 & -\ii\\
-\ii & 0\end{array}\right)\,.
\ee
Notice that these matrices satisfy $U^\dagger\sigma_1 U=\sigma_1$ and $\det U=1$, i.e.\  they are elements of $SU(1,1)$ (defined with respect to the $\mathbb{C}^2$ metric $\sigma_1$), which is known to be isomorphic to $SL(2,\mathbb{R})$. In other words,  $U(\mathfrak{t})$ and $U(\mathfrak{s})$ generate the $SU(1,1)$ representation of $SL(2,\mathbb{Z})$.

Using \eqref{excalt}, one can also check that the $2\times 2$ matrix $\calt^{IJ}(\phi)$ transforms as follows 
\be
\begin{aligned}
\calt(\phi+1)&=  U(\mathfrak{t})^{\dagger -1 } \calt(\phi) U(\mathfrak{t})^{-1} \, , \\
\calt\Big(-\frac1\phi\Big)&=U(\mathfrak{s})^{\dagger -1 } \calt(\phi) U(\mathfrak{s})^{-1} \, . 
\end{aligned}
\ee
Observing that the combination $\tilde{A}_{(3)I}-\bar\calg_{IK} A^K_{(3)}$ appearing in the boundary term \eqref{bdaction} transforms as $\calf_{(4)I}$, one can readily check that the bulk and boundary terms in the action \eqref{exbaction} are separately invariant under the $SL(2,\mathbb{Z})$ duality group. 
This shows that the $SL(2,\mathbb{Z})$ duality group is indeed a symmetry of the action \eqref{exbaction}.


\subsection{The mini-landscape of vacua}
\label{sec:landscape}

Let us study the vacua of the action \eqref{exbaction} (in the absence of membranes). As discussed in Section \ref{sec:equiv}, one can first integrate out the gauge three-forms by  picking up a particular symplectic constant vector 
\be\label{exconst}
\left(\begin{array}{c}
m^0 \\ m^1 \\ e_0 \\ e_1\end{array}\right)\in\mathbb{Z}^4
\ee
defined as in \eqref{incost} and rewriting \eqref{exbaction} in the form
\be\label{exbaction2}
S=-\int\d^4x\,e\,\left[\frac12 R+\frac{\del \phi\, \del\bar\phi}{4(\Im\phi)^2}+V(\phi,\bar\phi)\right],
\ee
where $V$ is a conventional $\caln=1$ potential 
\be\label{expot}
\begin{aligned}
V(\phi,\bar\phi)&= e^\calk\Big(K^{\phi\bar\phi}|D_\phi W|^2-3|W|^2\Big)\\
&=-\frac{e^{\hat K_0}}{2\Im\phi} \Big[(m^1)^2+(e_0)^2+4(m^0m^1+e_0e_1)\Im\phi 
		\\
		&\quad~~~~~~~~~~~~~~ +2(m^0e_0-m^1e_1)\Re\phi+((m^0)^2+(e_1)^2)|\phi|^2\Big]\,, 
\end{aligned}
\ee
with $\calk=K+\hat K_0$ and $W$  and $K$ are as in  \eqref{exSup} and \eqref{exKP}, respectively. Generically, the effective action \eqref{exbaction2} is not invariant under $SL(2,\mathbb{Z})$ transformations of $\phi$ (unless we  appropriately transform also the integration constants $m^I,e_J$). However, given the `microscopic' formulation with three-forms we started from, we can regard this breaking as {\em spontaneous} rather than explicit. 

Let us first consider the simplest  possibility:  $m^0=m^1=e_0=e_1=0$. In this case $W\equiv 0$ and hence $V\equiv 0$. So, we have a one-dimensional moduli space of vacua parametrized by an arbitrary expectation value of $\phi$. As standard in similar situations, one should identify two vacua related by an  $SL(2,\mathbb{Z})$  duality transformation. In view of the restriction \eqref{phipos}, the  moduli space of the inequivalent vacua can be identified with the familiar fundamental domain
\be\label{fundom}
\Big\{-\frac12\leq\Re\phi\leq \frac12\Big\}\cap \Big\{|\phi|\geq 1\Big\}\,.
\ee

On the other hand, this moduli space is drastically modified by any non-trivial set of constants \eqref{exconst}. It is useful to introduce the complex numbers
\be
\alpha_0\equiv e_0-\ii m^1\,,\quad \alpha_1\equiv e_1-\ii m^0
\ee
taking values in $\mathbb{Z}+\ii\mathbb{Z}$. Notice that the vector
\be
\vec\alpha\equiv  \left(\begin{array}{r}
\alpha_0\\
\alpha_1
\end{array}\right)
\ee
transforms as \eqref{calfvec} under the $SL(2,\mathbb{Z})$ duality tranformations, that is, in the fundamental $SU(1,1)$ representation generated by \eqref{USL(2,Z)}. The bosonic component of the superpotential \eqref{exSup} takes the form $W=\bar\alpha_0-\ii\bar\alpha_1\phi$ and the corresponding supersymmetric vacuum expectation value of $\phi$ (such that $D_\phi W|_{\phi_*}=0$) is 
\be\label{exvacua}
\phi_*=\ii\frac{\alpha_0}{\alpha_1}\,.
\ee
Taking into account \eqref{phipos}, we require
\be\label{exposcond}
\Im\phi_*= \frac{1}{|\alpha_1|^2}\Re(\alpha_0\bar\alpha_1) 
\ee
to be finite and positive.
In particular, this implies that the cases
$\alpha_1=0$ and $\Re(\alpha_0\bar\alpha_1)=0$ must be discarded.  
Then, given a certain $\vec\alpha$, \eqref{exvacua} is the only extremum of the potential \eqref{expot}. 

The condition $\Im\phi_*>0$ is equivalent to requiring that
\be\label{expos}
\Re(\alpha_0\bar\alpha_1)\equiv \frac12 \vec\alpha{}^\dagger \sigma_1\vec\alpha>0 . 
\ee
At the supersymmetric vacua \eqref{exvacua} the covariantly holomorphic superpotential $\calz$ takes the value
\be\label{calz*}
\calz_*=e^{\frac12\hat K_0}\frac{\bar\alpha_1}{|\alpha_1|}\sqrt{\Re(\alpha_0\bar\alpha_1)}
\ee
and the potential reduces to
\be
V_*=-3|\calz_*|^2=-3\,e^{\hat K_0}\Re(\alpha_0\bar\alpha_1)\,, 
\ee
which is strictly negative in view of \eqref{expos}, and thus determines the constant curvature of the AdS vacuum. 
The AdS radius (in natural units $M_{\rm P}=1$) is identified with the inverse of
\be\label{exmodZ}
|\calz_*|= e^{\frac12\hat K_0}\sqrt{\Re({\alpha_0\bar\alpha_1})}\, . 
\ee
Since $\alpha_0$ and $\alpha_1$ are integrally quantized, we should assume that $e^{\frac12\hat K_0}\ll 1$ in order  to be within the regime of reliability of our effective supergravity, which is equivalent to $|\calz_*|\ll 1$, therefore the AdS radius is much larger than the Planck length.

Now the $SL(2,\mathbb{Z})$ duality group  of the theory relates a vacuum \eqref{exvacua} associated with a certain set  of constants \eqref{exconst} (and a corresponding effective superpotential \eqref{exSup}) to another vacuum associated with a different set of constants and effective superpotential. 
It follows that, chosen a certain (generic) set of constants \eqref{exconst},  the domain of  $\phi$ is the entire upper half-plane \eqref{phipos}  (up to some possible residual and non-generic identification),  and not the fundamental domain  \eqref{fundom}. This is a purely four-dimensional realization of the flux-induced monodromy effects observed in string compactifications, see for instance \cite{Marchesano:2014mla} for a recent discussion.

\begin{figure}[H]
	\centering\includegraphics[width=7cm]{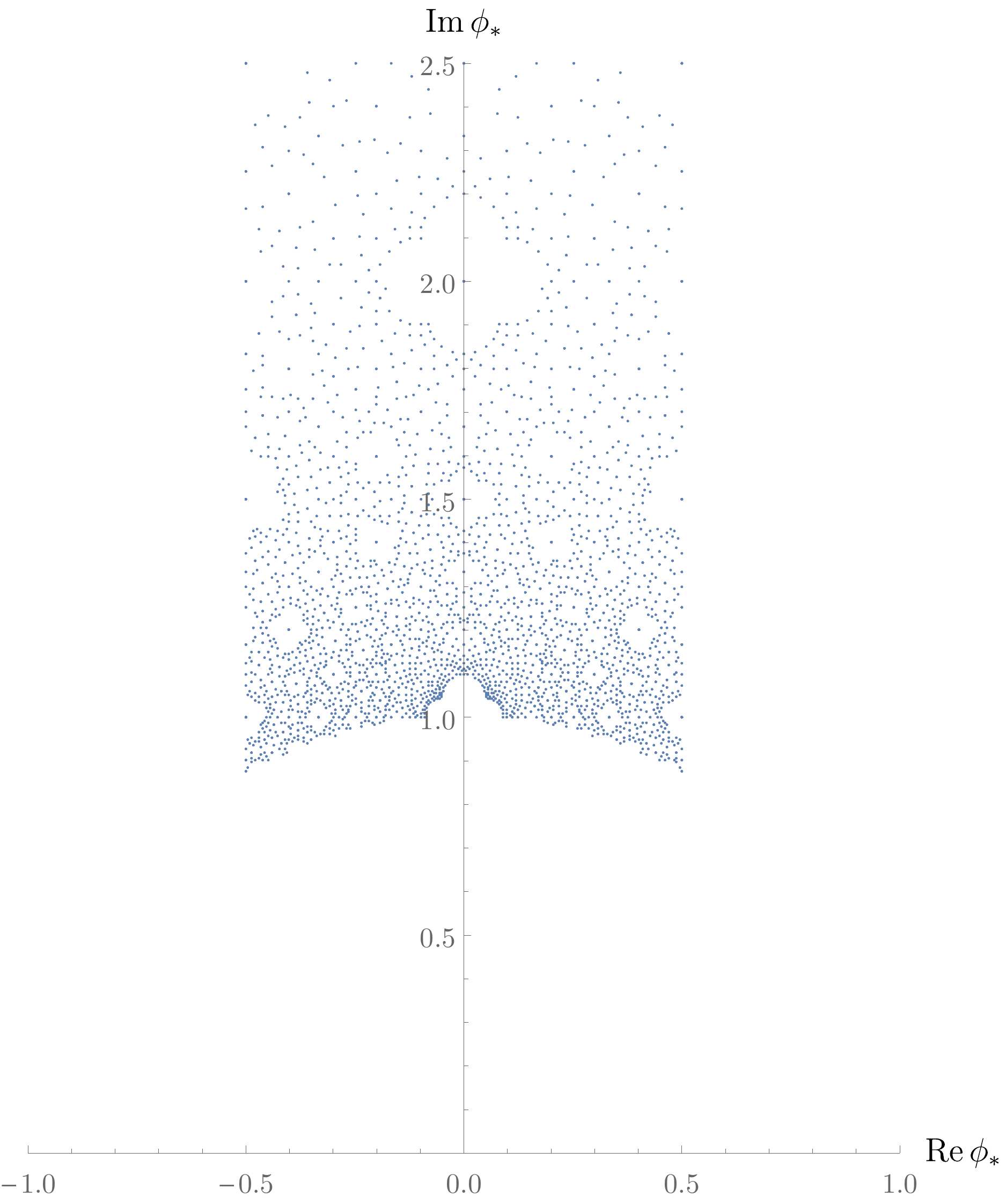}
	\caption{\footnotesize A sampling of vacua \eqref{exvacua} filling the fundamental domain \eqref{fundom}, for the values of the constants $e_I,m^J \in [-11,11]$.}
	\label{fig:vacua}
\end{figure}

On the other hand, in order to identify the {\em inequivalent} vacua, corresponding to inequivalent choices of the constants  \eqref{exconst} and of the corresponding effective potentials,  we can restrict ourselves to  the vacua \eqref{exvacua} which sit in the fundamental domain of \eqref{fundom}.
The set of such vacua is plotted in Fig.~\ref{fig:vacua}. A similar  set of vacua appears in the simplest models of type IIB flux compactifications on a rigid Calabi--Yau  \cite{Denef:2004ze}, in which $\phi$ can be identified with the  axion-dilaton. In the type IIB models one needs to impose the tadpole cancellation condition, which adds a constraint on the set of allowed vacua. In our formulation with gauge three-forms, the tadpole cancellation condition can be implemented as outlined at the end of Section \ref{sec:quant}.


\section{Domain walls  between aligned vacua}
\label{sec:exDWs}

In this section we explicitly construct a class of  domain walls of the kind discussed in Section \ref{sec:jumpingDW},  relating pairs of vacua $\phi|_{-\infty}=\phi_*$ and $\phi|_{+\infty}=\phi_*'$ of the form \eqref{exvacua}, corresponding to two sets of constants $\alpha_I$ and $\alpha_I'$ respectively. We make the simplifying assumption that the phases of $\calz_*$ and $\calz'_*$ are aligned  and that the phase of  $\calz(y)$ remains constant along the flow.
Of course, in order to have a (non-trivial) domain wall  $|\calz_*|$ and $|\calz'_*|$ should be different and then the corresponding vacua cannot be related by a $SL(2,\mathbb{Z})$ duality.
From \eqref{thetaflow} we see that we should impose $\Im(\dot\phi\,\del_\phi K)=0$ with $K$ as in \eqref{exKP}.  This is possible only if $\Re\phi$ is constant and equals to
\be\label{exaxion}
\Re\phi_*= -\frac{\Im(\alpha_0\bar\alpha_1)}{|\alpha_1|^2} . 
\ee
Clearly, we should also require $\Re\phi_*'=\Re\phi_*$\,.
Hence, 
\be
v(y) \equiv \Im \phi(y) 
\ee
is the only dynamical real field along the flow.  As in Section \ref{sec:jumpingDW}, we will assume that $|\calz|$ is always increasing along the flow, which drives the field $v$ from $v|_{-\infty}=v_*$ towards $v|_{+\infty}=v_*'$ and, at $y=0$, it crosses a membrane of charges $p^I,q_J$ such that
\be
q_0-\ii p^1=\alpha_0-\alpha_0'\,,\quad q_1-\ii p^0=\alpha_1-\alpha'_1\,.
\ee

The equations of the flow \eqref{floweq} are governed by the growth of $|\calz|$. In particular,   equation \eqref{susy2} reduces to
\be\label{flowv}
\dot{v} = 4v^2\, \frac{\d}{\d v} |\calz|\,.
\ee
For $y<0$, $|\calz|$ takes the following form
\be\label{Zv}
|\calz(v)|=\frac{e^{\frac12\hat K_0}}{2\sqrt{v}}\left[|\alpha_1|v+\frac{\Re(\alpha_0\bar\alpha_1)}{|\alpha_1|}\right]\,.
\ee
For $y>0$ the from of $|\calz|$ is obtained by replacing $\alpha_I$ with $\alpha_I'$ in \eqref{Zv}.

On the left of the membrane, $v_*$ is a global minimum of $|\calz|$. Hence, it is a repulsive fixed point of \eqref{flowv}, a flow is triggered and $v$ is driven away from $v_*$, letting the value of $|\calz|$ increase. When the membrane is reached at $y=0$, $v$ and consequently $|\calz|$ have evolved to certain values $v(0)$ and $|\calz|_{y=0}$. Here, the solution of the flow equations on the left should be glued to the one on the right. We are then led to impose the continuity of $v$ across $y=0$ while still keeping a growing $|\calz|$. However, since on the right of the membrane $v_*'$ is also a global minimum of $|\calz|$, $v_*'$ is a repulsive (rather than attractive) fixed point of \eqref{flowv}. Hence the solution to the flow equations is such that  $v$ reaches the value $v_*'$ at $y=0$ and then  remains constant
\be
v(y) = v_*' \quad \text{for}\;\,\,y\geq 0\,.
\ee
Correspondingly,  $|\calz|$ starts from $|\calz_*|$ at $y=-\infty$ and smoothly grows  until it reaches the membrane. At this point it jumps up to $|\calz'_*|$ and then remains constant (see Fig. \ref{fig:flowZ} for an example). Hence, on the right of the membrane, the background is just the AdS vacuum solution.

Recalling 
\eqref{DWtension2}, we  see that the bound
\be\label{DZbound}
T_{\rm DW} \geq   T_{\rm M}\,
\ee
is saturated if and only if on the left-hand side of the membrane $|\calz|$ is also constant. In the following we will first examine  the case in which the bound \eqref{DZbound} is saturated, leading to trivial flow equations on both sides of the membrane, and then we will consider an example for which the inequality \eqref{DZbound} strictly holds. 

As a warm up, 
let us assume that on the left of the membrane $\vec\alpha = 0$, so that the potential \eqref{expot} and $\calz$ are identically zero. Then, for $y<0$, the flow equations \eqref{floweq} are trivial and are immediately solved by taking $v$ and $D$ to be arbitrary constants. In particular, with no loss of generality, we can choose $D(y)\equiv0$ for $y<0$. 
Therefore, on the left of the membrane, the bulk is always at a fixed Minkowski vacuum. As discussed above, on the right of the membrane, the bulk is at its supersymmetric AdS vacuum, in which $v$ takes the constant value $v_*'$. Hence, by continuity, we should impose that $v(y)\equiv v_*'$ also for $y<0$.  Furthermore, by imposing also the continuity of the warp factor, we must set $D(y)=-|\calz'_*|y$ for $y>0$. 

It is worthwhile to mention that this particular case of trivial flow for both $y < 0$ and $y > 0$ can be realized only  when on the left-hand side the vacuum is Minkowski, owing to the freedom in choosing any constant value of $v$ for $y<0$.

\begin{figure}[H]
	\centering\includegraphics[width=6cm]{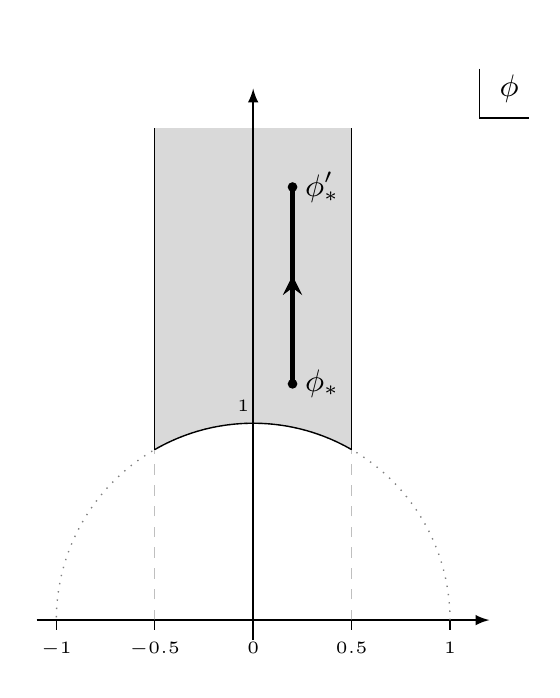}
	\caption{\footnotesize The fundamental domain  of $\phi$. Along the domain wall, $\phi$ flows up along the vertical line specified by $\Re \phi_*$.}
	\label{fig:FDf}
\end{figure}

Let us now consider a more involved example, for which the flow on the left side of the membrane is nontrivial. For any choice of initial constants $\alpha_0,\alpha_1$ and any $k\in\mathbb{Q}$ such that
\be
k\alpha_1\in\mathbb{Z}+\ii\mathbb{Z}
\ee
we can choose a jump to new constants
\be\label{constchoice}
\alpha_0'=\alpha_0+k\alpha_1\,,\quad \alpha_1'=\alpha_1 \, , 
\ee
which clearly satisfies $\Re\phi_*'=\Re\phi_*$. Notice that
\be
\Im\phi'_*=\Im\phi_*+k\,.
\ee
 The flow moves along the vertical direction of the upper-half-plane parametrized by $\phi$ (see Fig. \ref{fig:FDf}). With no loss of generality, we take $k>0$, so that $\Im\phi'_*>\Im\phi_*$.   From \eqref{calz*}, one can also see that $|\calz_*'|>|\calz_*|$ which is the default  assumption in Section \ref{sec:jumpingDW}. 

Under these restrictions, the initial and final values of $v(y)$ are
\be
v_*=\frac{\Re(\alpha_0\bar\alpha_1)}{|\alpha_1|^2}\,,\quad~~~~ v_*'=\frac{\Re(\alpha_0\bar\alpha_1)}{|\alpha_1|^2}+k 
\ee
and the membranes charges are
\be\label{exMcharges}
p^0=0\,,\quad p^1=k\,\Im\alpha_1\,,\quad q_0=-k\,\Re\alpha_1\,,\quad q_1=0\,.
\ee

\begin{figure}[ht!]
	\centering\includegraphics[width=9cm]{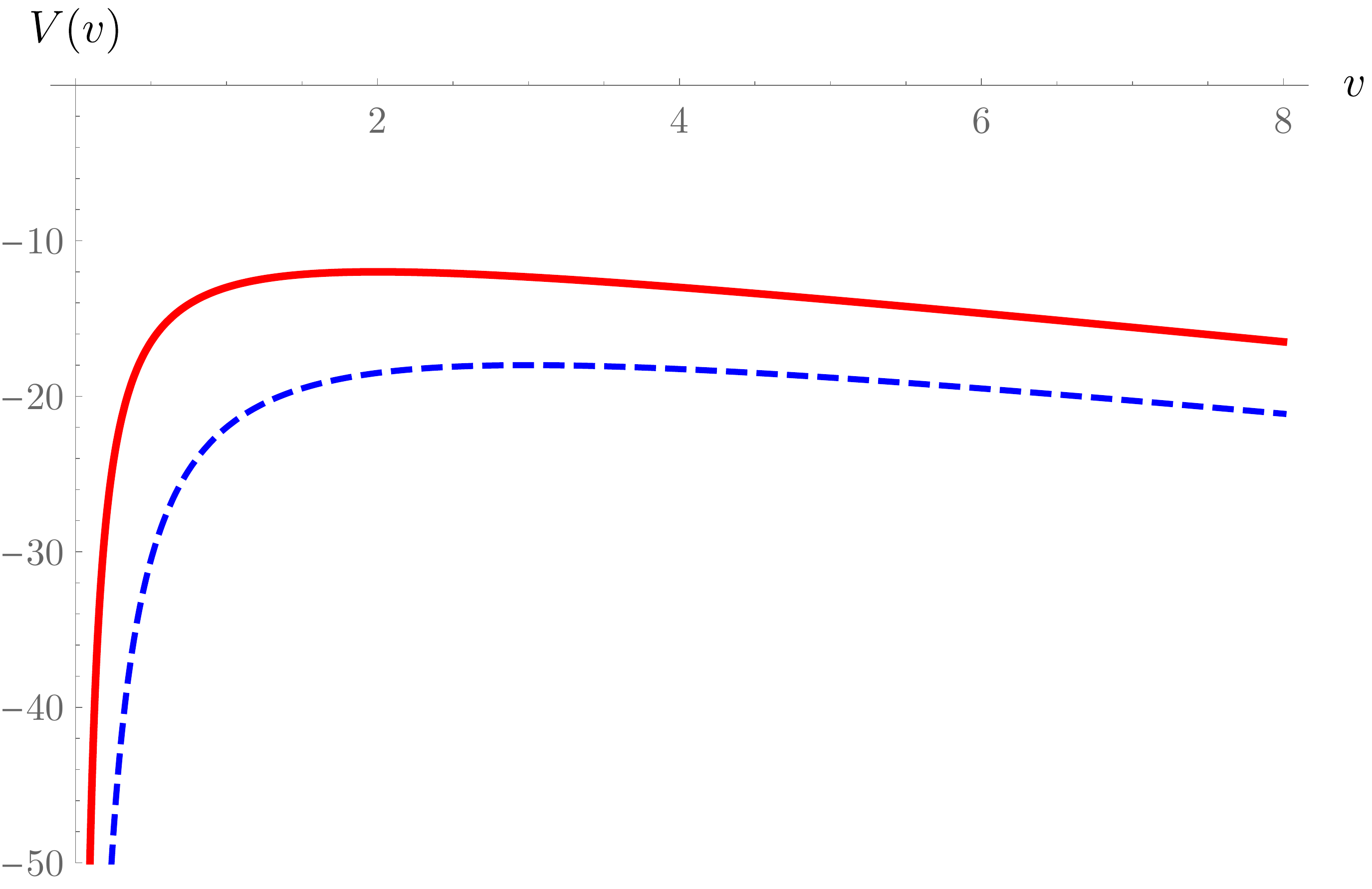}
	\caption{\footnotesize The potential \eqref{expot} for the choice of the constants $e_1 = m^0 =1$, $e_0 = m^1= 2$, $k=1$ and keeping $\Re\phi = 0$. The solid red line refers to the potential on the left of the membrane, while the dashed blue line to that on the right. This potential exhibits, on the left of the membrane, a supersymmetric AdS critical point located at $v_*=2$ and, on the right, a supersymmetric AdS critical point at $v_*'=3$.}
	\label{fig:Example_Pot}
\end{figure}

\begin{figure}[ht!]
	\centering\includegraphics[width=8cm]{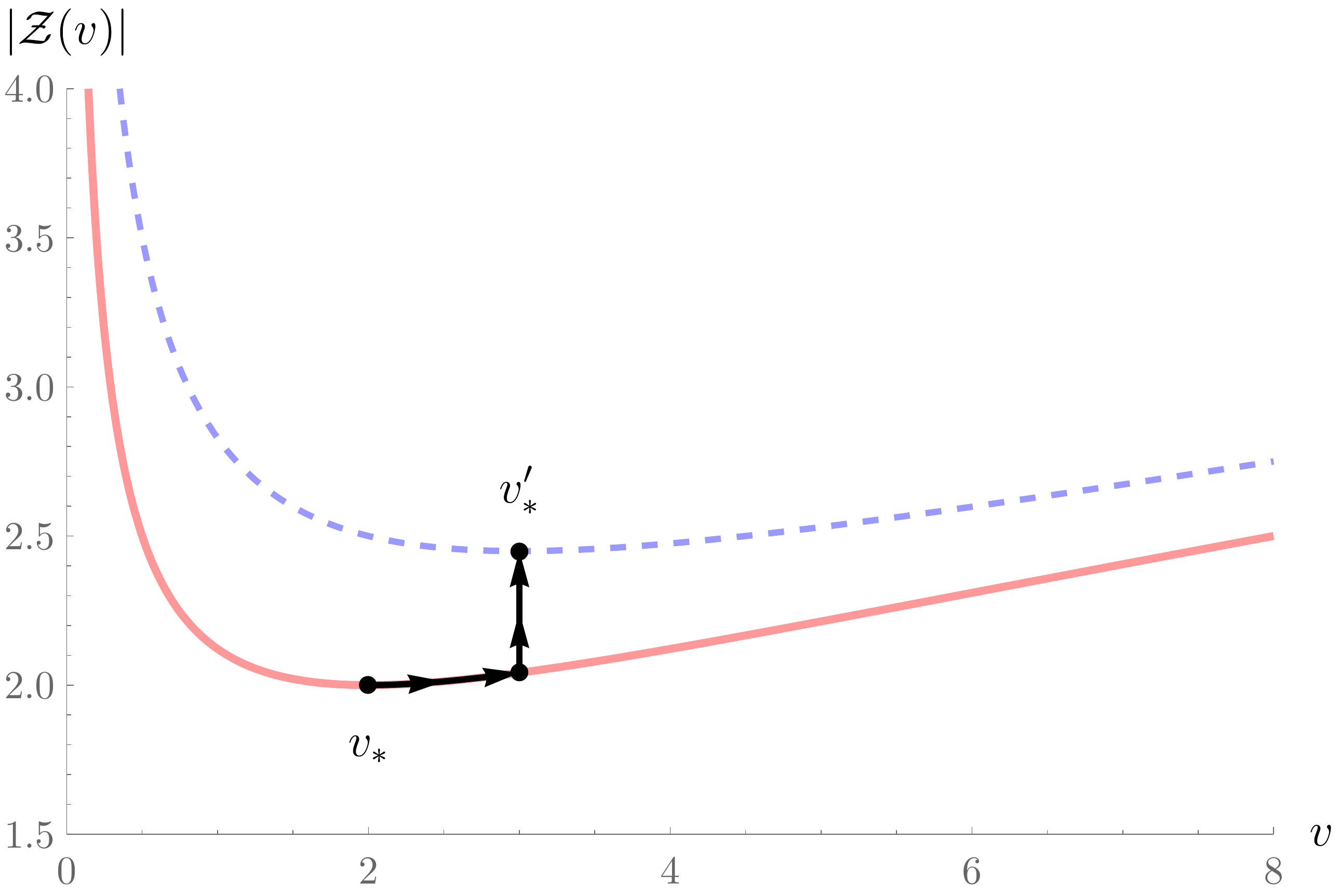}
	\caption{\footnotesize The flow of $|\calz|$ for the same set of parameters as in Fig. \ref{fig:Example_Pot}. The solid red line refers to $|\calz|$ on the left of the membrane, while the dashed blue line to that on the right. The flow drives $v_*$ towards the value $v_*'$, at which $|\calz|$ jumps so that $v$ is located at new supersymmetric vacuum on the right.}
	\label{fig:flowZ}
\end{figure}

We can now compute the function  $\calz(v,y)$ corresponding to our setting
\be
\calz(v,y)=\frac{\bar\alpha_1 e^{\frac12\hat K_0}}{2|\alpha_1|\sqrt{v}}\left[|\alpha_1| v+\frac{\Re(\alpha_0\bar\alpha_1)}{|\alpha_1|}+k|\alpha_1|\Theta(y)\right].
\ee
In agreement with \eqref{diffcalz}, $\calz(v,y)$ is discontinuous at $y=0$ and the width of the discontinuity is set by the tension of the membrane with the charges \eqref{exMcharges}
\be
\lim_{\varepsilon\rightarrow 0 }\Big|\calz(y_{\rm M}+\varepsilon)-\calz(y_{\rm M}-\varepsilon)\Big|=\frac{k|\alpha_1|e^{\frac12\hat K_0}}{2\sqrt{v|_{y=0}}}\equiv \frac12 T_M.
\ee
An example for the flow of $|\calz|$ is depicted in Fig. \ref{fig:flowZ}.

Consider now the flow equation \eqref{flowv}. For the examples under consideration, it takes the explicit form
\be
    \dot v=e^{\frac12\hat K_0}\sqrt{v}\left[|\alpha_1|v-\frac{\Re(\alpha_0\bar\alpha_1)}{|\alpha_1|}-k|\alpha_1|\Theta(y)\right]\label{exflow1}\, , 
\ee
which is solved by
\be
v(y)=\left\{\begin{array}{l}
v_*\coth^2\left[\frac12|\calz_*|(y+c)\right]\quad~~\text{for $y\leq 0$} \, ,
\\
v_*' \quad~~~~~~~~~~~~~~~~~~~~~~~~~~~~~\text{for $y\geq 0$} \, . 
\end{array}\right.\,
\ee
The integration constant $c$ must be negative, $c<0$, and  is fixed by the continuity at $y=0$, which imposes
$v_*\coth^2\left[\frac12c\,|\calz_*|\right]=v_*'
$
and   always admits a solution.

\begin{figure}[ht]
    \centering
	\includegraphics[width=7cm]{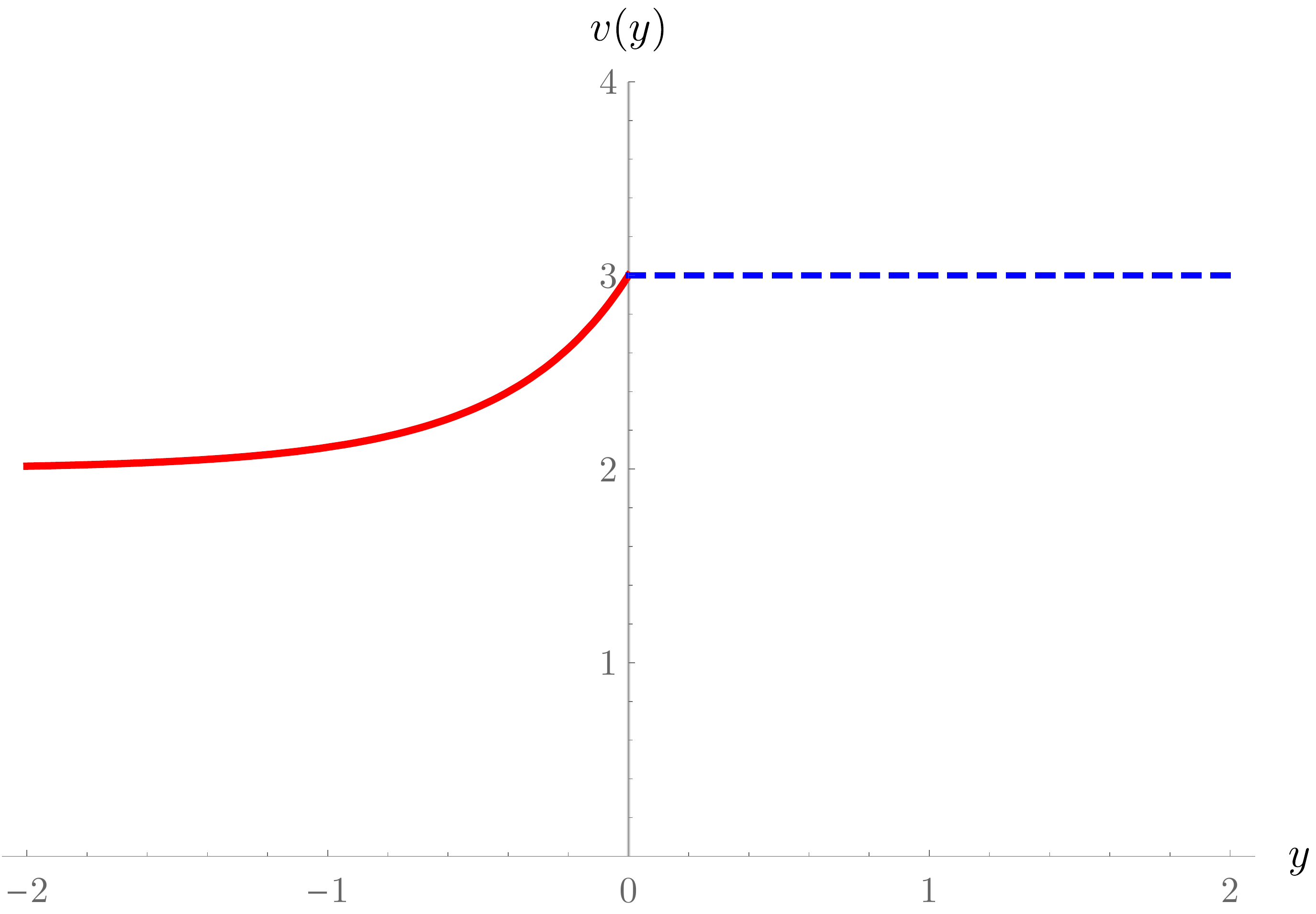} \includegraphics[width=7cm]{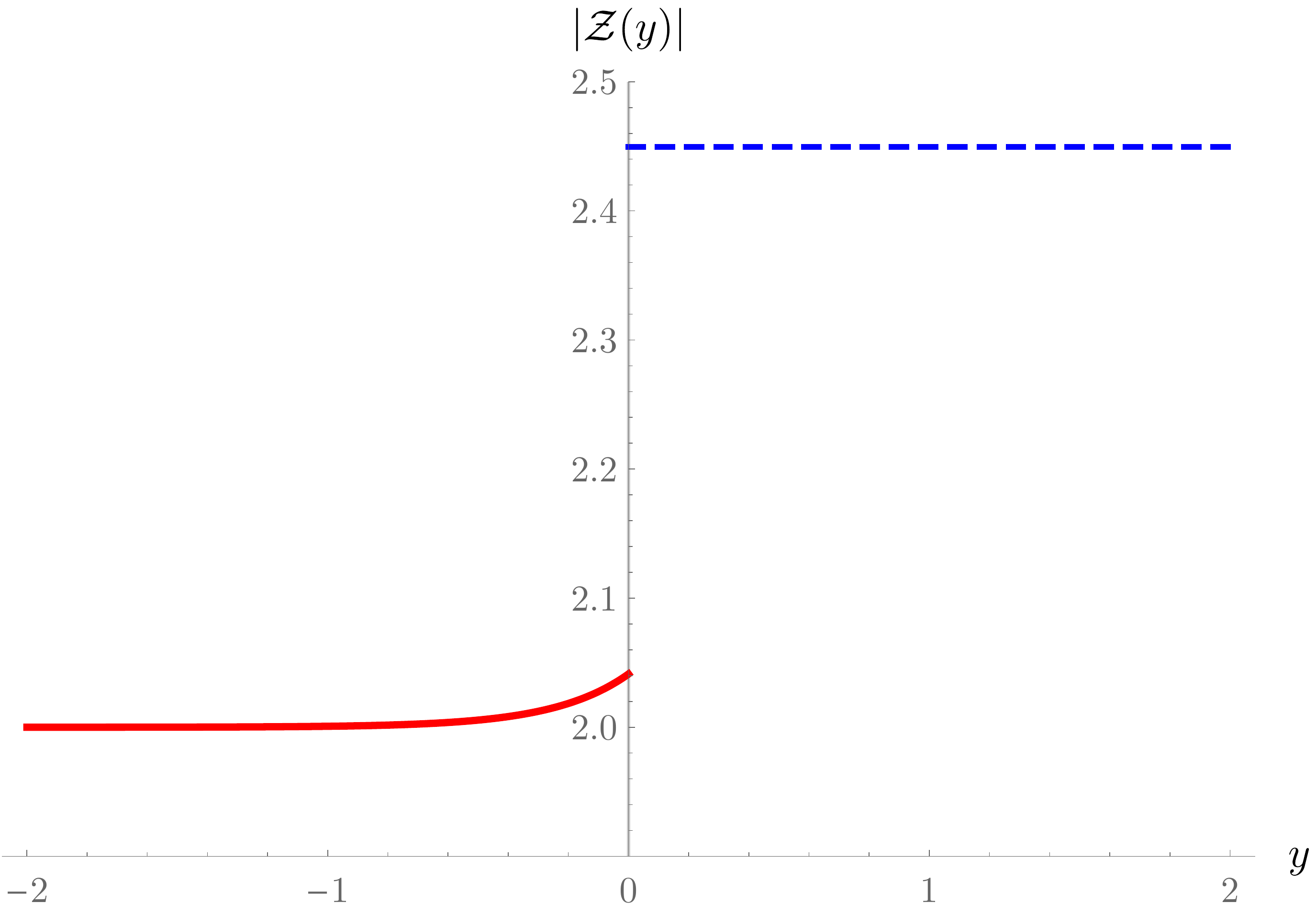}
	
	\centering
	\includegraphics[width=7cm]{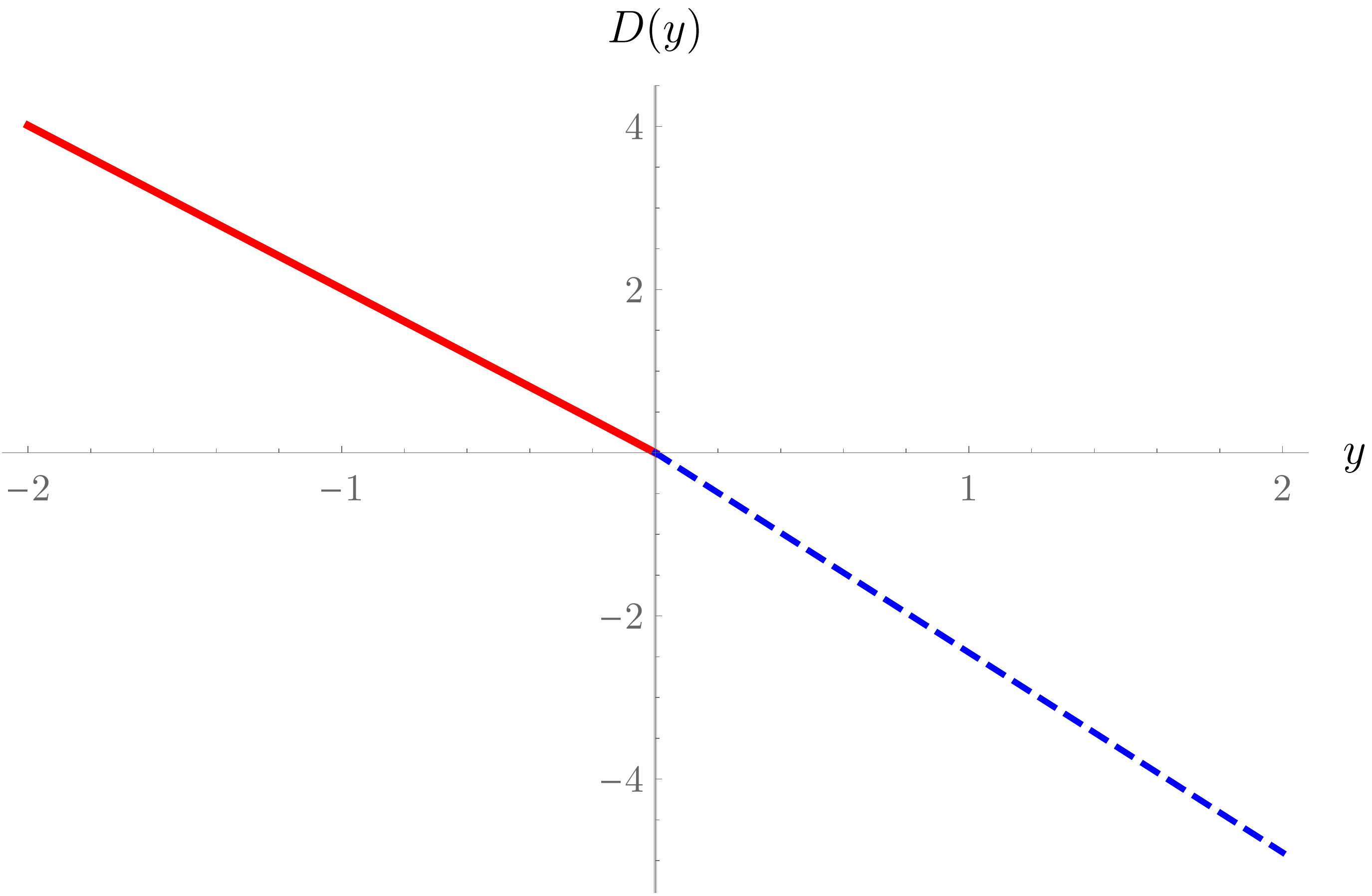} \includegraphics[width=7cm]{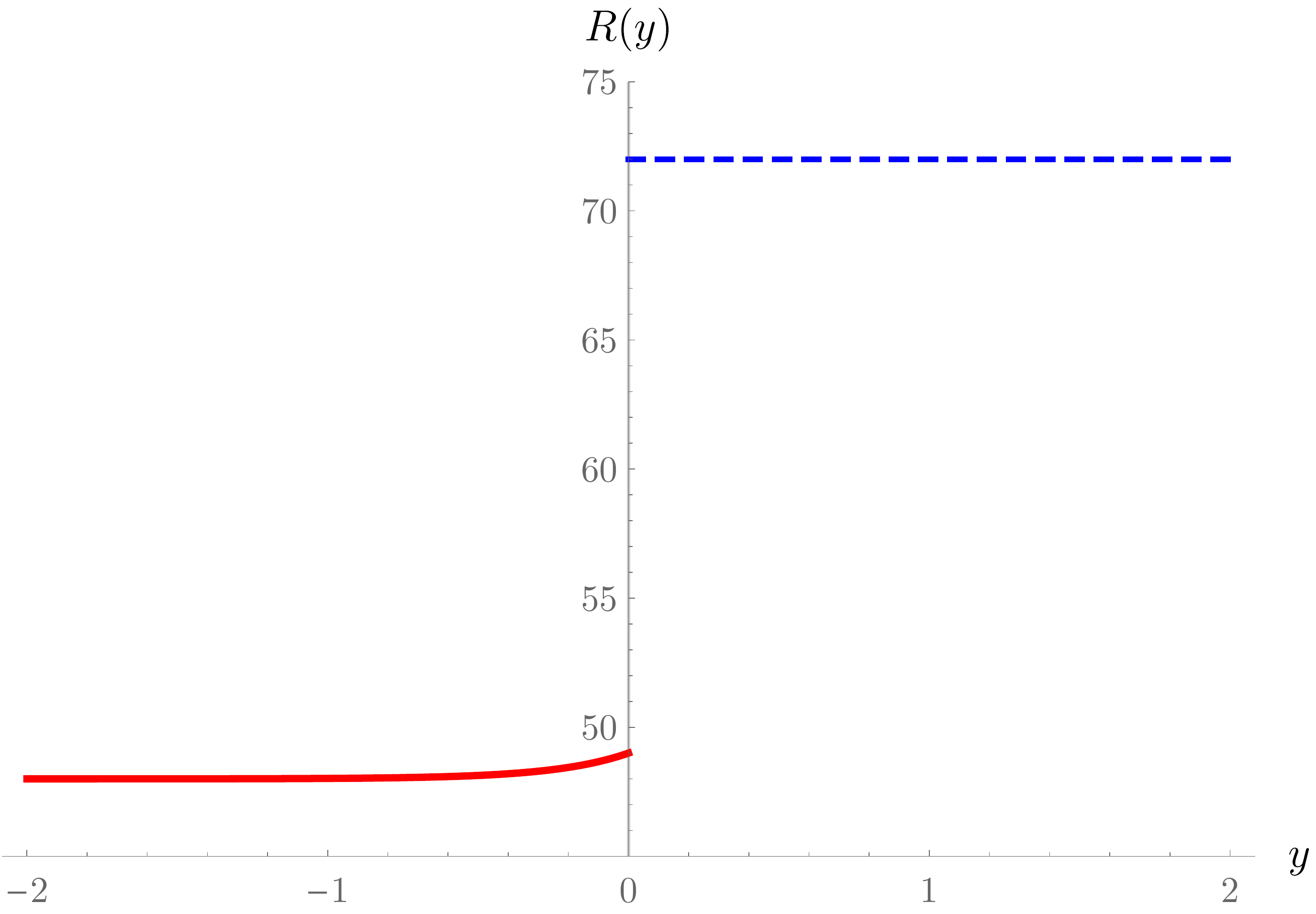}
	
	\caption{\footnotesize Above are depicted all the solutions to the flow equations \eqref{exflow1}, \eqref{exflow2} for the same set of parameters as in Fig.\,\ref{fig:Example_Pot}. The solid red lines refer to the quantities in the region on the left of the membrane, while the dashed blue lines to those on the right. On the top left there is the evolution of the scalar field: starting from the critical point on the left of the membrane, the field $v$ is driven towards the critical point on the right of the membrane. On  the  top  right  there  is  the modulus of the covariantly  holomorphic superpotential $|\calz|$, which is always increasing. On the bottom left there is the warping $D(y)$, which is always decreasing, using which the curvature, on the bottom right, can be obtained. As expected from the AdS vacua, the curvature is at a fixed positive value when the field $v$ reaches the vacua and, even though not explicitly shown here, is singular at the point $y=0$. In the figures $y$ is $|\calz_*|^{-1}$ units.}
	\label{fig:Memb_Simple_Example_ClassI}
\end{figure}
Since the bulk exhibits a non-trivial flow only on the left-hand side of the membrane, the membrane tension is given by the vacuum expectation value of $\phi_*'$ to the right of the membrane 
\be
T_{\rm M}=\frac{k|\alpha_1|^2e^{\hat K_0}}{|\calz'_*|}.
\ee

 We still have to solve the equation for the warping  \eqref{susy1}, which in the present case reads
 \be
    \dot D=-\frac{e^{\frac12\hat K_0}}{2\sqrt{v}}\left[|\alpha_1|v+\frac{\Re(\alpha_0\bar\alpha_1)}{|\alpha_1|}+k|\alpha_1|\Theta(y)\right]\label{exflow2}.
\ee
 It also admits an analytic solution given by 
 \be
 D(y)=\left\{\begin{array}{l}
 d+{e^{-\frac12\hat K_0}\left(\log(-\sinh u)+\log\cosh u\right)}\quad~\text{for $y\leq 0$}\,,\\
  - |\calz_*'| y\quad~~~~~~~~~~~~~~~~~~~~~~~~~~~~~~~~~~~~~~~~~\text{for $y\geq 0$}\,, 
 \end{array}\right.
 \ee
where we have set to zero an arbitrary additive constant  and   $u(y)\equiv\frac12 |\calz_*|(y+c)$. The integration constant $d$  is fixed by imposing the continuity of $D(y)$ at $y=0$.

A couple of final comments. Notice that on the left-hand side of the membrane  the deviation of the complete solution from the  AdS vacuum   is concentrated within a length of the  same order of the  AdS radius $|\calz_*|^{-1}$. Hence, in this sense, the domain wall may be considered as `thick'. Furthermore, clearly, we can make a coordinate redefinition $y\rightarrow y-y_{\rm M}$ to get a solution with the membrane localised  at any point $y_{\rm M}$.

\section{Conclusions}

In this paper we have studied and expanded the $\caln=1$ supergravities including double three-form multiplets  introduced  in \cite{Farakos:2017jme}. We have focused on the subclass of models in which the dynamics of the double three-form multiplet sector is governed by a special K\"ahler structure and is covariant under symplectic tranformations.

Into this setup we have included supermembranes of arbirary (quantised) charges, which naturally couple to the supersymmetric completion of the three-form potentials via a WZ term. Given the WZ term, the worldvolume $\kappa$-symmetry of the membrane action fixes the form of its NG term which includes the dependence on the bulk scalar sector in the way expected from string compactification models (see Appendix \ref{app:kappa} for the proof of $\kappa$-symmetry and Appendix \ref{app:gener} for further generalizations).

The back-reaction of the membrane induces a jump in the vevs of the four-form field-strengths. Hence, from a more conventional supergravity perspective (which can be retrieved from the three-form theory by setting the field-strengths on-shell), this implies the appearance of an effective superpotential with different coupling constants on the left and on the right-hand side of the membrane. 
Within this setup we have examined how supersymmetric vacua corresponding to different four-form flux integration constants separated by the membrane are connected by `jumping' BPS domain walls. 
As a simple and instructive example we have considered a model with two double three-form multiplets and found explicit analytic solutions describing jumping domain walls therein.
Thus, our results generalize the class of the BPS domain walls of four-dimensional $\mathcal N=1$ supergravities studied previously e.g. in \cite{Cvetic:1992bf,Cvetic:1992st,Cvetic:1992sf, Cvetic:1993xe,Cvetic:1996vr,Ovrut:1997ur,Ceresole:2006iq,Huebscher:2009bp}. 

We believe that the results of this paper provide an appropriate starting point for describing, from an effective four-dimensional perspective, non-trivial dynamical processes involving at the same time membranes, fluxes and the scalar sector of flux compactifications, as for instance those considered in \cite{Bousso:2000xa,Feng:2000if}. In particular, in this paper we have only considered the effects of membranes on flat BPS domain walls, postponing  the study of other possible dynamical effects (for example, the nucleation of non-BPS membrane bubbles) and their physical implications to the future.


\subsection*{Acknowledgements}
We thank  G.~Dall'Agata, I.~García-Etxebarria, S. Kuzenko and I.~Valenzuela for useful discussions. Work of I.B. was supported in part by the
Spanish MINECO/FEDER (ERDF) EU  grant FPA 2015-66793-P, by the Basque Government Grant IT-979-16, and the Basque Country University program UFI 11/55. 
Work of F.F. is supported in part by the Interuniversity Attraction Poles Programme 
initiated by the Belgian Science Policy (P7/37), and in part by support from the KU Leuven C1 grant ZKD1118 C16/16/005. 
Work of S.L.\ and L.M.\ was partially supported by the Padua University Project CPDA144437. 
Work of D.S. was supported in part by the Russian Science Foundation grant 14-42-00047 in association with Lebedev Physical Institute  and by the Australian Research Council project No. DP160103633. D.S. is grateful to the Department of Physics, UWA and the School of Mathematics, the University of Melbourne for hospitality at an intermediate stage of this project.

\begin{appendix}

\section{Super-Weyl transformations}
\label{app:superspace}

The Lagrangian \eqref{WIactio} is invariant under the super-Weyl transformations of the chiral superfields and the super-vielbein \cite{Howe:1978km, Siegel:1978fc}
\be
\begin{aligned}
	Z &\rightarrow e^{-6\Upsilon} Z , \\
	E^a_M &\rightarrow e^{\Upsilon+\bar\Upsilon} E^a_M , \\
	E^\alpha_M &\rightarrow e^{2\Upsilon-\bar\Upsilon}\left(E^\alpha_M - \frac{\ii}{2} E_M^\alpha \sigma_a^{\alpha\dot\alpha} \bar\cald_{\dot\alpha} \bar\Upsilon \right) . 
\end{aligned}
\ee
After singling out the chiral compensator as in \eqref{SYPHI}, we can think of the super-Weyl transformation as acting on the chiral compensator only
\be
	Y \rightarrow e^{-6\Upsilon} Y
\ee
leaving the chiral superfields $\Phi^i$ invariant. Under a general K\"ahler transformation
\be
\label{KTransf}
\begin{aligned}
K(\Phi, \bar\Phi) \rightarrow K(\Phi, \bar\Phi) + f(\Phi) + \bar f (\bar\Phi)
\end{aligned}
\ee
the Lagrangian \eqref{WIactio} is not invariant. Such invariance is only restored if \eqref{KTransf} is accompanied by a super-Weyl rescaling of the compensator and the superpotential, namely 
\be
\begin{aligned}
	Y &\rightarrow e^{-f(\Phi)} Y \,, \\
	W(\Phi) &\rightarrow e^{-f(\Phi)} W(\Phi)\,.
\end{aligned}
\ee
In other words, the chiral compensator and the superpotential are holomorphic sections of a complex line bundle over the K\"ahler manifold.

The prepotential $\mathcal P$ which determines the structure of the three-form gauge superfield $\mathcal A_3$ in \eqref{super3form} transforms under the Weyl rescaling as follows
\be\label{WP}
\mathcal P \to e^{-2(\Upsilon+\bar\Upsilon)}\mathcal P.
\ee
This ensures that $\mathcal A_3$ is Weyl invariant \cite{Kuzenko:2017vil}.

\section{General bosonic action}
\label{app:boslagr}
With the choice of the kinetic function as in \eqref{Omega} and \eqref{Omega0}, the most general superfield action built from \eqref{3formlagr} leads to the bosonic component action of the following form
\be
\begin{aligned}
	S_{\rm bos}=&- \int \d^4x\,e\, \Big(\frac{1}{2}R+ G_{IJ} f^I{}_i \bar{f}^{J}{}_{\bar\jmath}\; \del \phi^i \del\bar{\phi}^{\bar \jmath}  + \hat K_{p \bar q}\,\del t^p\del\bar t^{\bar q} \Big)
	\\
	&+S_{\text{3-forms}} +S_{\hat W}\,
	\end{aligned}
\ee
where the three-form action $S_{\text{3-forms}}$ is
\be
 S_{\text{3-forms}} = \int \d^4x\,e\, \calt^{IJ}{}^*\!\bar\calf_{4I}\, {}^*{}\!\calf_{4J}+ S_{\rm 3-forms,\, bd}
\ee
and the $\hat W$-depending action $S_{\rm \hat W}$ is (where we use the property of the superpotential $\hat W = \hat{W}_K f^K$) 
\be
\begin{aligned}
	S_{\hat W} =& \int \d^4x\,e\, \left\{ - \text{e}^\calk 
	\left[ \hat{K}^{\bar q p}
	- \frac{1}{\gamma} \hat{K}^{\bar q l} \hat{K}_l \hat{K}^{\bar l p} \hat{K}_{\bar l} \right]  
	\hat{W}_p \overline{\hat{W}}_{\bar q} \right\} 
	\\
	&+ \Re \int \d^4x\,e\, \Bigg\{- \frac{\ii}{\gamma (f {\cal M} \bar f)}  \hat{K}_{\bar q} \hat{K}^{\bar q p}\hat{W}_p \bar f^I {}^*{}\!\calf_{4I} +
	\\
	&
	+\ii ( f {\cal M} \bar f) \left[ \hat{W}_K G_{IL} 
	{\cal M}^{LK} {\cal M}^{IN} - \frac{\hat{W} \bar f^N}{(f {\cal M} \bar f)^2} \right] \, {}^*{}\!\calf_{4N} 
	\Bigg\} + S_{\hat W,\, \rm bd}\,.
\end{aligned}
\ee
Here $\gamma$ is defined as in \eqref{noscale}. 
The boundary terms are given by
\be
\begin{aligned}
    S_{\rm 3-forms,\, bd} &= \, 2\Re\int_\calb \calt^{IJ}(\tilde A_{3I}-\calg_{IK}A_3^K){}^*\!\calf_{4J}\, 
\end{aligned}
\ee
and 
\be
\begin{aligned}
	S_{\hat W,\, \rm bd} =&\, \Re \int_{\calb} \Bigg\{- \frac{\ii}{\gamma (f {\cal M} \bar f)}  \hat{K}_{\bar q} \hat{K}^{\bar q p}\hat{W}_p \bar f^I (\tilde A_{3I} - \calg_{IK} A_3^K) +
	\\
	&
	+\ii ( f {\cal M} \bar f) \left[ \hat{W}_K G_{IL} 
	{\cal M}^{LK} {\cal M}^{IN} - \frac{\hat{W} \bar f^N}{(f {\cal M} \bar f)^2} \right] (\tilde A_{3N} - \calg_{NP} A_3^P)
	\Bigg\}\,.
\end{aligned}
\ee

\section{Supersymmetry transformations of fermions}
\label{app:susy}
In the double three-form supergravity under consideration the supersymmetry transformations of the gravitino and the chiralini, in the bosonic background,  have the following form 
 \be\label{susy}
 \begin{aligned}
	\delta \psi_m{}^\alpha &= -2 \hat{\cald}_m \zeta^\alpha -\ii  e_m{}^c e^{-\frac{\calk}{2}} W (\epsilon \sigma_c \bar{\zeta})^\alpha\,,\\
	\delta \chi^i_\alpha &= \sqrt{2} \zeta_\alpha e^{\frac{\calk}{2}} K^{\bar \jmath i} ( \overline{W}_{\bar \jmath}+K_{\bar \jmath}\overline{W}) -\ii \sqrt{2} \sigma_{\alpha\dot{\beta}}{}^a \bar{\zeta}^{\dot{\beta}} \del_a \phi^i\, ,\\
	\delta \rho^p_\alpha &= \sqrt{2} \zeta_\alpha e^{\frac{\calk}{2}}\hat{K}^{\bar q p} K_{\bar q} \overline{W} -\ii \sqrt{2} \sigma_{\alpha\dot{\beta}}{}^a \bar{\zeta}^{\dot{\beta}} \del_a t^p\,,
 \end{aligned}
 \ee
 where $W$ and $W_i$ were defined in \eqref{W=}, 
the covariant derivative of the supersymmetry parameter is given by 
 \be
 \begin{split}
	\hat{\cald}_m \zeta^\alpha &\equiv \del_m \zeta^\alpha + \zeta^\beta \omega_{m\beta}{}^\alpha - \frac{\ii}{2} \cala_m \zeta^\alpha \, ,
\end{split}
\ee
and the $U(1)$ K\"ahler connection is 
 \be
	\cala_m = \frac{\ii}{2} \left(K_i \del_m \phi^i - K_{\ibar} \del_m \bar{\phi}^\ibar+\hat{K}_p \del_m t^p - \hat{K}_{\bar p} \del_m \bar{t}^{\bar p} \right)\,. 
\ee

In the absence of $T_p$ multiplets, the BPS condition on the domain wall ansatz discussed in Section \ref{sec:jumpingDW} is obtained by setting to zero the corresponding variations \eqref{susy}, which reduce   to
\be\label{susydw}
 \begin{aligned}
	\delta \psi_y{}^\alpha = & -2 \dot\zeta^\alpha +\ii \cala_y  \zeta^\alpha-\ii e^{\frac{\calk}{2}} W (\epsilon \sigma_{\underline{y}} \bar{\zeta})^\alpha\,,\\
	\delta \psi_i{}^\alpha = & e^D \left[ \dot{D}\ (\zeta \sigma_{\underline y}\bar{\sigma}_{\iund})^\alpha - \ii e^{\frac{\calk}{2}} W (\epsilon \sigma_{\iund} \bar{\zeta})^\alpha\right]\,,\\
	\delta \chi^i_\alpha = &\sqrt{2} \zeta_\alpha e^{\frac{\calk}{2}} K^{\bar \jmath i} ( \overline{W}_{\bar \jmath}+K_{\bar \jmath}\overline{W})  -\ii \sqrt{2} \sigma_{\alpha\dot{\beta}}{}^{\underline y} \bar{\zeta}^{\dot{\beta}} \dot \phi^i\, .
 \end{aligned}
 \ee

\section{Proof of  $\kappa$-symmetry}
\label{app:kappa}

In superspace, we mostly follow notation and conventions of \cite{Wess:1992cp}.  In particular, for the superspace superform algebra of this and the following appendices,  we adopt the inverse-index notation and the external-derivative  acts from the right. \footnote{Here and in the following appendices, the results of \cite{Bandos:2010yy,Bandos:2011fw,Bandos:2012gz,Bandos:2002bx,Bandos:2003zk} are employed. To pass from the (mostly minus) notation used there to that of \cite{Wess:1992cp}, one should change the sign of the metric, $\eta^{ab}\mapsto -\eta^{ab}$, the spin connection $\omega_a{}^b\mapsto -\omega_a{}^b$, curvature 
$R_a{}^b\mapsto -R_a{}^b$ and of the right-handed fermionic covariant derivative,  $\bar{{\cal D}}_{\dot\alpha}\mapsto -\bar{{\cal D}}_{\dot\alpha}$, rescale 
the chiral superfield of supergravity ${\cal R} \mapsto {\cal R}/8$, 
and assume that the following quantities do not change the sign: $E^a, \sigma^a_{\alpha\dot\alpha}, \varepsilon_{abcd},  \varepsilon^{ijk}, G^a$.}

The action of the supermembrane in the background of supergravity and three-form multiplets, \eqref{complmembrane}, \eqref{susyNG} and \eqref{susyWZ},  can be written in the following form 
\be\label{Smembr} S_M= S_{NG}+S_{WZ}= -2\int_{\calc} \d^3\xi \sqrt{-\det\, h} \; | T| + \int_{\calc}  {\cal A}_3\; , \qquad 
\ee
where $ T$ is a composite special chiral superfield 
\be\label{cZqp}  T = q_I S^I- p^I {\cal G}_I(S)= 
q_IS^I -p^I{\cal G}_{IJ}(S)S^J \, , 
\ee
which is constructed as 
\be\label{cZqp=bD2P}  T = -\frac \ii4 (\bar{{\cald}}^2-8 \calr) \cal P 
\ee
from the composite prepotential 
\be\label{WZprepot1} \calp = q_I \calp^I- p^I \tilde{{\cal P}}_I= 
-2 q_I\calm^{IJ}\Im\Sigma_J +2p^I\Im(\bar\calg_{IJ}\calm^{JK}\Sigma_K) 
\ee
in which $\calp^I$ and $\tilde{{\cal P}}_I$ were defined in \eqref{realprepot1}. 
The three-form  ${\cal A}_3$ in the WZ term of \eqref{Smembr} is constructed as in \eqref{super3form} with the composite prepotential  \eqref{WZprepot1}. 
The field strength of this super-three-form is
\begin{eqnarray} \label{bH4=SGZ}   \calh_{4}   &=& \d{\cal A}_3= \;   {\ii} E^b\wedge E^a \wedge \bar E^{\dot\alpha} \wedge \bar  E^{\dot\beta }\bar{\sigma}_{ab\; \dot{\alpha}\dot{\beta}} T - {\ii} E^b\wedge E^a \wedge E^\alpha \wedge E^\beta \sigma_{ab\; \alpha\beta}\bar{ T}  \qquad \nonumber \\  && - \frac{\ii}6  E^c\wedge E^b\wedge E^a \wedge \bar E^{\dot\alpha} \epsilon_{abcd} \sigma^d_{\alpha\dot\alpha} {\cal D}^{\alpha}{ T}  -\frac{\ii}6 E^c\wedge E^b\wedge E^a \wedge E^\alpha \epsilon_{abcd} \sigma^d_{\alpha\dot\alpha} \bar{{\cal D}}^{\dot\alpha}\bar{ T} \nonumber \\  && + \frac{1}{96} E^{d} \wedge E^c \wedge E^b \wedge E^a \epsilon_{abcd} \left(({\cal D}{\cal D}-24\bar{\calr})
{ T}+(\bar{{\cal D}}\bar{{\cal D}}-24\calr) 
\bar{ T} \right) 
\,. \qquad
\end{eqnarray}

The measure $\d^3\xi$ in the Nambu-Goto type term is defined by 
\be\label{d3xi=} \d\xi^i\wedge \d\xi^j\wedge \d\xi^k=\epsilon^{ijk}\,\d^3\xi \; . \ee
This  implies  the identities 
\be\label{d3xi=ids}
\d^3\xi  \sqrt{-h}= \frac 1 3 {}^{*_3}\!E_a\wedge E^a\; , \qquad \d^3\xi\,  \delta \sqrt{-h}=  {}^{*_3}\!E_a\wedge \delta E^a\; , \qquad 
\ee
where the action of worldvolume Hodge duality operation ${}^{*_3}$ on a one-form is defined by 
\be\label{*3} 
{}^{*_3}\!E^A:= \frac 1 2 \d\xi^j\wedge \d\xi^i \sqrt{-h} \epsilon_{ijk}h^{kl} E_l^A
\; . \qquad 
\ee
This latter can be used to write the  variation of the Nambu-Goto action with respect to the embedding coordinates $z^M(\xi)$ in the form 
\be\label{vNG} 
\delta S_{NG}= - 2\int_{\calc} {}^{*_3}\!E_a\wedge \delta E^a\, |T| -  2 \int_{\calc} \d^3\xi\sqrt{-h}
\frac {T \delta\bar{T} +  \delta T\, \bar{T} } {|T |}
\; , \qquad 
\ee
while the variation of the Wess-Zumino term is \footnote{In the case of a closed  membrane or an infinitely extended membrane (with a proper behaviour at infinity)  the total derivative term 
does not contribute.}
\be\label{vWZ} 
\delta S_{WZ}= \int_{\calc} \delta \cala_3 =\int_{\calc}(i_{\delta z}\d\cala_3+ \d\, i_{\delta z} \cala_3)=\int_{\calc} i_{\delta z} \calh_4. 
\ee
Here for varying ${\calc}$ 
we used the Lie derivative formula
$ \delta_z = \d i_{\delta z} + i_{\delta_z} \d$ and its Lorentz covariant extension
\be 
\delta_z = {\cal D}  i_{\delta z} + i_{\delta z} {\cal D} \, , 
\ee
which is equivalent to the Lie derivative modulo local Lorentz transformations. 

We are searching for $\kappa$-symmetry transformations, leaving the supermembrane action invariant, in the  form which is common for a general class of superbranes, i.e.
\be
\label{kappa:=}
\begin{split}
&i_\kappa E^a =\delta_\kappa z^M E_M^a(z)=0\; , \qquad  \\
&i_\kappa E^\alpha =\delta_\kappa z^M E_M^\alpha (z)=\kappa^\alpha\; ,\qquad i_\kappa E^{\dot\alpha} =\delta_\kappa z^M E_M^{\dot\alpha} (z)=\bar{\kappa}{}^{\dot\alpha}\; . \qquad 
\end{split}
\ee
For this transformations the variations of the bosonic supervielbein, chiral superfields  and the WZ term take the form
\bea\label{varkapEa=}
 &\delta_\kappa E^a = {\cal D}i_\kappa E^a+ i_\kappa {\cal D}E^a = -2\ii E^\alpha 
 (\sigma^a\bar{\kappa})_\alpha  + 2\ii  ({\kappa}\sigma^a)_{\dot\alpha} E^{\dot\alpha}  \; , \\
 \label{varkap-ch=}
 &\delta_\kappa  T =\kappa^\alpha {\cal D}_\alpha T\; , \qquad \delta_\kappa  \bar T =\kappa^\alpha \bar{{\cal D}}_{\dot\alpha} \bar T\, , 
 \eea
\be\label{ifcalh4=} 
\begin{split}
i_\kappa \calh_4 = \;  &- 2\ii E^b\wedge E^a \wedge \bar E^{\dot\alpha} (\bar{\kappa}\bar{\sigma}_{ab})_{\dot{\alpha}} T +2\ii E^b\wedge E^a \wedge E^\alpha  (\sigma_{ab}\kappa)_{\alpha}\bar{ T}- \\  
& - \frac{\ii}6 E^c\wedge E^b\wedge E^a  \epsilon_{abcd} (\sigma^d\bar{\kappa})_{\alpha} {\cal D}^{\alpha}{ T}  -\frac{\ii}6 E^c\wedge E^b\wedge E^a \epsilon_{abcd} (\kappa\sigma^d)_{\dot\alpha} \bar{{\cal D}}^{\dot\alpha}\bar{ T} 
\; .
\end{split}
\ee
Now, using the identities
\be
\begin{aligned}
&E^b\wedge E^c\wedge E^\beta \sigma_{bc\, \beta}{}^\alpha = -2\, {}^{*_3}\!E^a\wedge E^\beta (\sigma_a\,{\Gamma})_{\beta}{}^\alpha \; , \\
&\d^3\xi \sqrt{-h} {\Gamma} = 
\frac \ii {3!}  \sigma^a\epsilon_{abcd}
E^b\wedge E^c\wedge E^d  \, , 
\end{aligned}
\ee
one can check that the variation of the WZ term (\ref{vWZ}) cancel 
the variation (\ref{vNG}) of the NG term, provided 
\be\label{kk}
\kappa_\alpha =\frac T {|T|} ({\Gamma}\bar{\kappa})_\alpha\; .
\ee
This is exactly the condition  (\ref{kappaproj}) of the main text.


\section{Generic systems of 3-form matter, supergravity and supermembranes}
\label{app:gener}

The supermembrane interaction with a single three-form multiplet is described by the equations from the previous section, if we consider the 
special chiral superfield 
$T$ to be fundamental, i.e. expressed through a single {\em fundamental} real prepotential $\calp$ rather than composite as in (\ref{cZqp}). In this case the chiral superfield $T$ has the auxiliary field $F_T=F+\ii{}^*\!F_4$ whose real part is a scalar and the imaginary part is the dual of the single four-form.

Now, to describe general systems of supergravity and three-forms coupled to the membrane
we introduce a set of chiral superfields of conformal weight 3, $ Z^{\Lambda}$ ($\Lambda =(\mathfrak I,I)$), where the indices $\mathfrak I$ and $I$ label the subsets of double and single three-form superfields. In this set the conformal compensator $Y$ can be chosen at will. It can be either single- or double three-form superfield. Then the other superfields  are associated with the double or single 3-form matter supermultiplets. Note that there also is a third case in which the conformal compensator is not among the independent fields of the set $Z^{\Lambda}$ coupled to the membrane. Then the supermembrane couples to supergravity only via the {\em physical} three-form superfields. In this case the off-shell supergravity can be consistently chosen to be the conventional old-minimal supergravity with the both components of its complex auxiliary field being scalars (and not three-forms).

The general action for the supermebrane coupled to the superfields $ Z^{\Lambda}$ has the following form 
  \begin{eqnarray}\label{Sp2+chiral=1}
  S_{p=2}= -2\int_{\calc} \d^3 \xi \sqrt{{ -}h} |q_{\Lambda} {Z}^{\Lambda}| +  q_{\mathfrak I} \int_{\calc} {C}_3{}^{\mathfrak I} +  \bar{q}_{\mathfrak I} \int_{\calc} {\bar{C}}{}_3^{\mathfrak I} +
  q_I  \int_{\calc} {A}{}_3^{I} \; ,\qquad
\end{eqnarray}
where $C_3^{\mathfrak I}$ are the complex super three-forms associated with the double three-form supermultiplets and the real super three-forms $A_3^{ I}$ are associated with the single three-form ones. 

The  action is invariant under the kappa-symmetry transformations (\ref{varkapEa=}) and \eqref{kk} in which $T={{q}_{\Lambda}{ Z}^{ \Lambda}}$.


\section{Membrane equations of motion}
\label{app:mem-eq}

Here we enlist the equations of motion coming from the complete action \eqref{boscomp}+\eqref{compmembrane} and their form after employing the domain wall ansatz \eqref{ds2=ansatz}.

First, the equation of motion of the graviton is
 \be\label{graveom}
R^{mn} - \half g^{mn} R = T^{mn} +  g^{mn} V +\int\, \d^3 \xi \,T_M\frac{\sqrt{-h}}{\sqrt{-g}} \delta\left(x^m-z^m(\xi)\right) h^{ab} \partial_a z^{m} \partial_b z^{n}
 \ee
 with
 \be
 \begin{aligned}
  T^{mn} &=
  K_{i\bar \jmath} \left(g^{mn} g^{pq} -2g^{mp} g^{nq}\right)\del_p \phi^i \del_q \bar\phi^j.
 \end{aligned}
 \ee
Taking the trace of \eqref{graveom}, we get the following equation for the scalar curvature 
 \be\label{scalarR}
 -\frac{R}{2} = K_{i\bar \jmath}\, \del^m \phi^i \del_m \bar\phi^j  +2 V + \frac 32 \int\d^3 \xi\,\frac{\sqrt{-h}}{\sqrt{-g}} \delta\left(x^m-z^m( \xi)\right) T_M.
 \ee
In the static gauge $x^\mu(\xi)= \xi^i\delta_i^\mu$, in which the only nontrivial worldvolume bosonic field is $y(x)$,  the last term in (\ref{scalarR}) reduces to
 $\frac 3 2 T_M\frac{\sqrt{-h}}{\sqrt{-g}}  \delta\left(y-y(x)\right)$ and the equation for the scalar curvature becomes 
 \be\label{scalarR=}
 -\frac {R} 2 = K_{i\bar \jmath}\, \del^m \phi^i \del_m \bar\phi^j  +2 \calt^{IJ} {}^*\!\bar\calf_{4I}\, {}^*\!\calf_{4J} + \frac 3 2 \frac{\sqrt{-h}}{\sqrt{-g}} \delta\left(y-y(x)\right) T_M.
 \ee
The domain wall ansatz \eqref{ds2=ansatz} implies that 
the only nonvanishing (vielbein) component of the spin connection is
\be
\omega_i^{a3}=-\delta_i^ae^{2D(y)}\dot{D}\,,
\ee
so that the curvature two-form reduces to 
\be
\label{Rab=DW}
	R^{ab} = -e^a\wedge e^b \dot D^2+2\delta_3^{[a} e^{b]}\wedge dy\, \ddot D\; 
\ee
and the Ricci scalar is
\be
\label{DWR}
	R = 6\left( 2\dot D^2+\ddot D\right)\; . 
\ee
We can then combine \eqref{scalarR} with the flow equation \eqref{susy1}, immediately getting
\be
\label{diffcalz2}
\frac{\d |\calz|}{\d y}= K_{i\bar\jmath}\dot\phi^i \dot{\bar\phi}^{\bar\jmath} +\frac12 T_{\rm M}\,\delta(y)
\ee
coherently with \eqref{diffcalz}. The above equation is also implied by the three-form field equations \eqref{fluxeq2} if one uses the definition of $W$ in \eqref{W=}.
This shows the consistency of the supergravity equations of motion with the domain wall ansatz \cite{Ovrut:1997ur} and the flow equations \cite{Huebscher:2009bp}.

Let us now come to the equations of motion for the membrane. When all the fermions are set to zero, in the Einstein frame the supermembrane action has the form \eqref{compmembrane}, which we write as
\be\label{SM}
S_M = -2 \int\limits_{\calc} d^3\xi \sqrt{-h} T_M + q^I\int\limits_{\calc} A_3^I - p^I \int\limits_{\calc} \tilde{A}_{3I}
\ee
with 
\be\label{TM=} T_M= 
{\cal Z}= e^{\frac 1 2 \calk}| (q_If^I(\phi)- p^I\calg_I(\phi))|\, .
\ee
The supermembrane equations are then 
\be\label{mem-eq=Gen}
{\cal D}_i\left(\sqrt{-h}h^{ij}E_{ja}T_M \right)= 
\sqrt{-h} \,  {\cal D}_a T_M  + \frac 1 {3!}\varepsilon_{abcd}\varepsilon^{ijk}E_i^bE_j^cE_k^d\left(q_I {}^*\!F^I_4-p^I\!{}^*\tilde{F}_{I4}\right)\; ,
\ee
where 
$E_i^a$ are coefficients of the pull back of the bosonic vielbein, $E_i^a=\partial_i x^m e_m^a(x(\xi))$. In the bosonic background \eqref{ds2=ansatz}, after fixing the `static gauge' $x^\mu=\xi^i \delta_i^\mu$, these latter acquire the form 
\be\label{Eia=static}
E_i^a= \partial_i y(x) \delta_3^a + e^{D(y)}  \delta_i^a
\ee
so that 
\be\label{h-dwall}
\begin{split}
h_{ij} &=  e^{2D(y)}  \eta_{ij}+ \partial_i y(x)\partial_j y(x)\; ,\\
h^{ij} &= e^{-2D(y)} \left( \eta_{ij}- \frac  { e^{-2D(y)}  \partial^i y(x)\partial^j y(x)}{1 +e^{-2D(y)}  \partial^k y(x)\partial_k y(x) } \,\right)\; ,
\end{split}
\ee
where $\partial^i y(x)=\eta^{ij}\partial_j y(x)$.
Let us consider a ground state solution of these equations in the domain wall background \eqref{ds2=ansatz} in which the scalar fields depend only on the transverse coordinate $y$,
$T_M=T_M(y)$, the three-form gauge potentials have the form \eqref{3form=DW}, 
while  
\be\label{*F=ans}
{}^*\!F_4^I= -e^{-3D} \frac \d {\d y} \alpha^I(y) \; , \qquad {}^*\!\tilde{F}_{4I}= -e^{-3D} \frac \d {\d y} \tilde\alpha_I(y)\;  , 
\ee
and ${\cal D}_a T_M = \delta_a^3 \frac \d {\d y} T_M$. 

It is natural to assume that the ground state solution describes a  flat membrane worldvolume such that
\be\nonumber
\partial_iy=0\quad \Rightarrow \qquad y(x)=y_0={\rm const}\; ,
\ee
where $y_0$ indicates the place of the membrane in the bulk.

This implies that $E_i^a= e^{D(y_0)}\delta_i^a$, $h_{ij}=e^{2D(y_0)}\eta_{ij}$, ${\cal D}_i\left(\sqrt{-h}h^{ij}E_{ja}T_M \right)=-3\delta_a^3\dot{D}e^{D}T_M(y_0) $, and the only nontrivial ($a=3$) component of the equations (\ref{mem-eq=Gen}) takes the form 
\be\label{memem}
\left.\left(3\dot{D}  T_M+ \frac \d {\d y}T_M - q_I{}^*\!F^4_I(y)+p^I{}^*\!\tilde{F}_{4I}(y)\right)\right|_{y=y_0}=0 \, , 
\ee
or, taking into account (\ref{*F=ans}),  
we have 
\be\label{memEqAp}
{}  \left.   \frac \d {\d y}\left( e^{3D} T_M  + q_I\alpha^I-p^I\tilde\alpha_I \right)\right|_{y=y_0} =0\; ,
\ee
where $T(y_0)=T_M$ is the membrane tension.

Let us now connect the previous discussion with that of Section 6. There, we considered the (super)membrane action \eqref{compmembrane} as part of the action for the interacting system including dynamical supergravity and matter fields.  In general, actions of this kind possess the {\em bulk} diffeomorphism invariance which can be used  to choose, directly in the action, the gauge in which the embedding of the supermembrane worldvolume into the bulk is described by the equation 
\be\label{y=0} 
y(x)=0\,,
\ee 
so that the transverse fluctuations of the membrane look `frozen'. Nevertheless,  the supermembrane equations can still be obtained from the action in this gauge. They appear as self-consistency conditions of the supergravity (and matter) field equations (see  \cite{Bandos:2001jx,Bandos:2005ww} and \cite{Bandos:2003tm} for  discussion and more references). A particular manifestation of this effect is that the membrane equations \eqref{memeq} are satisfied identically due to 
the  consequence \eqref{bulkid} of the flow equations \eqref{floweq1}.

\end{appendix}

\providecommand{\href}[2]{#2}\begingroup\raggedright\endgroup

\end{document}